\documentclass[]{aastex63}

\received{October 3, 2022}
\revised{November 4, 2022}
\accepted{November 16, 2022}
\submitjournal{AJ}
\shorttitle{The open clusters Berkeley 68 and Stock 20}
\shortauthors{Yontan}
\usepackage{xcolor}
\usepackage{rotating}
\begin{document}

\title{An investigation of open clusters Berkeley 68 and Stock 20 using CCD UBV and Gaia DR3 data}

\correspondingauthor{Talar Yontan}
\email{talar.yontan@istanbul.edu.tr}

\author[0000-0002-5657-6194]{Talar Yontan}
\affiliation{Istanbul University, Faculty of Science, Department of Astronomy and Space Sciences, 34119, Beyaz\i t, Istanbul, Turkey}

\begin{abstract}
\noindent

We performed detailed photometric and astrometric analyses of the open star clusters Berkeley 68 and Stock 20. This was based on ground based CCD {\it UBV} photometric data complemented by space-based {\it Gaia} Data Release 3 (DR3) photometry and astrometry. 198 stars were identified as likely cluster members for Berkeley 68 and 51 for Stock 20. Two-color diagrams were used to derive the reddening and photometric metallicity for each cluster. The reddening for Berkeley 68 is $E(B-V)=0.520 \pm 0.032$ and $0.400 \pm 0.048$ mag for Stock 20. Photometric metallicity [Fe/H] is  $-0.13 \pm 0.08$ dex for Berkeley 68, and $-0.01 \pm 0.06$ dex for Stock 20. Keeping as constant reddening and metallicity, we determined the distance moduli and ages of the clusters through fitting isochrones to the {\it UBV} and {\it Gaia} based color-magnitude diagrams. Photometric distances are $d=3003 \pm 165$ pc for Berkeley 69 and $2911 \pm 216$ pc for Stock 20. The cluster ages are $2.4 \pm 0.2$ Gyr and $50 \pm 10$ Myr for Berkeley 68 and Stock 20, respectively. Present day mass function slopes were found to be $\Gamma = 1.38 \pm 0.71$ and $\Gamma = 1.53 \pm 0.39$ for Berkeley 68 and Stock 20, respectively. These values are compatible with the value of \citet{Salpeter55}. The relaxation times were estimated as 32.55 Myr and 23.17 Myr for Berkeley 68 and Stock 20, respectively. These times are less than the estimated cluster ages, indicating that both clusters are dynamically relaxed. Orbit integration was carried out only for Berkeley 68 since radial velocity data was not available for Stock 20.  Analysis indicated that Berkeley 68 was born outside the solar circle and belongs to the thin-disc component of the Milky Way.

\end{abstract}

\section{Introduction}
\label{sec:introduction}
Compared to globular clusters, open clusters (OCs) are relatively more young and more metal-rich stellar systems which are located toward the Galactic plane. OCs are groupings of stars with similar physical properties. The number of member stars in OCs ranges from a dozen to thousands.  These stars form at the same time through the collapse of a molecular cloud \citep{Mckee07}. Initially these stars are gravitationally bound into the cluster, and hence share common positional and kinematic characteristics. The ages, distances, and chemical compositions of cluster member stars are similar, however the stellar masses are different \citep{Maurya20}. The study of OC stars through astrometric and photometric observations allows determination of astrometric and astrophysical parameters unavailable to the study of single stars \citep{Dias21}, partially explaining the interest in these systems. Such analysis helps in understanding the formation, evolution, and dynamics of stars \citep{Lada03, Portegies10}. Moreover, utilizing astrophysical parameters (such as distance, metallicity, and age) together with kinematic properties makes OCs good tracers for investigating formation and evolution of the Galactic disc \citep{Cantat-Gaudin20, He21, Hou21}. 

Identification of cluster member stars is negatively affected by the presence of field stars. Precise determination of cluster parameters requires the accurate separation of cluster members from such field stars. {\it Gaia} data releases \citep{Gaia16, Gaia18, Gaia21, Gaia22} have made available astrometric data of unprecedented accuracy\footnote{The uncertainties in {\it Gaia} DR3 are 0.01-0.02 mas for $G\leq15$ mag, and reach about 1 mas at $G=21$ mag.  The uncertainties of trigonometric parallax ($\varpi$) are 0.02-0.07 mas for $G\leq17$ mag, 0.5 mas for $G=20$, and reach 1.3 mas for at $G=21$ mag. For sources with $G\leq17$ mag the proper motion uncertainties are 0.02~--~0.07 mas yr$^{-1}$, reaching 0.5 mas yr$^{-1}$ at $G=20$, and 1.4 mas yr$^{-1}$ at $G=21$ mag. For sources within $G\leq17$ mag the $G$-band photometric uncertainties are 0.3~--~1 mmag, increasing to 6 mmag at $G=20$ mag.}, allowing such separations to be made more accurately \citep{Monteiro19, Cantat-Gaudin18}.  Many recent studies have made use of {\it Gaia} data in membership analyses of stars in the direction of open clusters and to investigate cluster properties \citep{Cantat-Gaudin18, Cantat-Gaudin19, Castro-Ginard18, Castro-Ginard19, Bisht19, Bisht20, Liu19}. The third {\it Gaia} data release (DR3) was made public on 2022 June 13$^{\rm rd}$. It contains data on more than 1.5 billion sources, and provides both {\it Gaia} early data release 3 (EDR3) and full {\it Gaia} DR3 data that span a period of 34 months (between 25 July 2014 and 28 May 2017).  The database provides combined radial velocities for about 33.8 million objects down to $G<14$ mag which is two magnitudes fainter than the second  {\it Gaia} data release (DR2) \citep{Gaia18}. The median formal precision of the radial velocities are 1.3 km s$^{-1}$ at $G=12$ and 6.4 km s$^{-1}$ at $G=14$ mag \citep{Katz22}.

In this paper we investigate two open clusters, Berkeley 68 and Stock 20, which are located in the second Galactic quadrant. Since the locations of both clusters are very close to the Galactic plane, they are strongly affected by field stars.  Statistical removal of such field contamination is an important part of the project. This study is a continuation of an ongoing research project investigating previously poorly analyzed open clusters in our Galaxy \citep{Yontan15, Yontan19, Yontan21, Yontan22,  Ak16, Banks20, Akbulut21, Koc22}. It aims to determine the main parameters of Berkeley 68 and Stock 20 using CCD {\it UBV} data together with the most recent database of the {\it Gaia} mission.  

%---------------------------------------------------------------

Berkeley 68 ($\alpha=04^{\rm h} 44^{\rm m} 13^{\rm s}$, $\delta= +42^{\rm o} 08^{\rm '} 02^{\rm''}$, $l=162^{\rm o}.\!04$, $b=-2^{\rm o}.40$) was classified by \citet{Ruprecht66} as Trumpler type IV2p with a low central stellar concentration. Analysing 2MASS and PPMXL data, \citet{Kharchenko12} obtained color excesses, age, and proper motion components for the cluster as $E(B-V)=0.67$ mag, $\log t=9.11$ yr, and ($\mu_{\alpha}\cos\delta$, $\mu_{\delta}$)=(-5.860, 0.300) mas yr$^{-1}$ respectively. \citet{Cantat-Gaudin18} listed member stars for 1,229 open clusters and derived their physical parameters considering {\it Gaia} DR2 photometric and astrometric data. For Berkeley 68 they identified 247 members with a membership probability greater than 0.5. The cluster distance was estimated as $d = 3223_{-786}^{+1531}$ pc and the proper motion components as ($\mu_{\alpha}\cos\delta$, $\mu_{\delta}) = (2.310 \pm 0.012, -1.312 \pm 0.008$) mas yr$^{-1}$. \citet{Liu19} employed {\it Gaia} DR2 data within an automatic isochrone fitting scheme, estimating the cluster distance, age, and the proper motion as $d=3559 \pm 608$ pc, $t=1.45 \pm 0.09$ Gyr, and ($\mu_{\alpha}\cos\delta$, $\mu_{\delta}) \; = \; (2.321\pm0.227, -1.322\pm0.203$) mas yr$^{-1}$. The first detailed CCD $U\!BV\!R_{\rm c}I_{\rm c}$ photometric study was performed by \citet{Maurya20}. They combined these observations with the 2MASS $JHK_{\rm S}$ near-IR data \citep{Cutri03} and {\it Gaia} DR2 astrometry \citep{Gaia18}. \citet{Maurya20} estimated the core radius of Berkeley 68 to be $ r_{\rm c}=0.6 \pm 0.3$ arcmin through fitting the radial density profile of \citet{King62}. They also identified 264 member stars based on {\it Gaia} DR2 proper motions. \citet{Maurya20} fitted the zero-age main sequence isochrone from \citet{Schmidt82} to the $(U-B)\times (B-V)$ two-color diagram (TCD), deriving values for the $E(U-B)$ and $E(B-V)$ color excesses. The reddening vector was estimated as $E(U-B)/E(B-V)=0.60 \pm 0.03$ and the color excess as $E(B-V)=0.52 \pm 0.04$ mag. The cluster age and distance moduli were estimated via joint consideration of the  $(B-V) \times V$ and $(V-I) \times V$ color-magnitude diagrams (CMDs). Fitting the  theoretical isochrones of solar metallicity $Z=0.0152$ presented by \citet{Marigo17} to the CMDs led to the age being calculated as $\log t= 9.25 \pm 0.05$ yr and the de-reddened distance modulus as $\mu_{\rm V}=13.7 \pm 0.2$ mag. This  corresponds to the isochrone distance of the cluster being $d=2554 \pm 387$ pc. \citet{Cantat-Gaudin20} found the age, isochrone distance, and interstellar reddening of Berkeley 68 as $t=2.1$ Gyr, $d=3383$ pc, and $A_{\rm V}=1.38$ mag (which corresponds to a color excess of $E(B-V)=0.445$ mag) respectively. 

Stock 20 ($\alpha=00^{\rm h} 25^{\rm m} 16^{\rm s}$, $\delta= +62^{\rm o} 37^{\rm '} 26^{\rm''}$, $l=119^{\rm o}\!\!.93$, $b=-00^{\rm o}\!\!.10$) was classified by \citet{Ruprecht66} as a Trumpler type II2p with medium richness. \citet{Kharchenko12} presented a catalogue of astrophysical data for 642 galactic open clusters including Stock 20. They considered 2MASS and PPMXL data, applying homogeneous methods and algorithms to obtain the color excess, distance, age, and mean proper motion components of Stock 20 as $E(B-V)=0.2$ mag, $d=1100$ pc, and $\log t=8.337$ yr, and ($\mu_{\alpha}\cos\delta$, $\mu_{\delta}$)=(-2.56, -3.24) mas yr$^{-1}$ respectively. \citet{Buckner13} employed a decontamination procedure to $J\!H\!K_{\rm S}$ photometry of 378 known OCs (including Stock 20) so as to estimate the density of stars across the cluster bodies, together with a galactic model to determine distances without the use of isochrone fitting. Using this method \citet{Buckner13} found the distance of the Stock 20 as $d=2600$ pc. \citet{Cantat-Gaudin18} identified 50 stars in the direction of Stock 20 as cluster members, assuming a membership probability estimated as being greater than 0.5, and subsequently calculated the cluster distance as  $d=2692_{-571}^{+990}$ pc together with the proper motion components as ($\mu_{\alpha}\cos\delta$, $\mu_{\delta}$)=($-3.237\pm0.010, -1.124\pm0.011$) mas yr$^{-1}$. \citet{Liu19} estimated the distance, age, and proper motion of the cluster as $d=2890\pm 309$ pc, $t=8\pm 1$ Myr, and ($\mu_{\alpha}\cos\delta$, $\mu_{\delta}$)=($-3.164\pm0.258, -0.950\pm0.308$) mas yr$^{-1}$, respectively. \citet{Cantat-Gaudin20} found the age, isochrone distance, and interstellar reddening of Stock 20 to be $t=19$ Myr, $d=2598$ pc and $A_{\rm V}=1.09$ mag (which corresponds to a color excess of $E(B-V)=0.352$ mag), respectively. Also, the fundamental parameters given in the literature are listed in Table \ref{tab:literature} for both clusters.

In the current study we determined the membership probability of cluster stars, mean proper motions, and distances of the OCs Berkeley 68 and Stock 20 using with ground-based {\it UBV} photometric observations combined with high-precision astrometry and photometry taken from the {\it Gaia} DR3 database. We present fundamental parameters, luminosity and mass functions, the dynamical state of mass segregation, and kinematic and galactic orbital parameters of the two clusters.

% Table 1
\begin{table*}
\setlength{\tabcolsep}{1.4pt}
\renewcommand{\arraystretch}{0.80}
%\small
  \centering
  \caption{Fundamental parameters for Berkeley 68 and Stock 20 derived in this study and compiled from the literature: Color excesses ($E(B-V$)), distance moduli ($\mu$), distances ($d$), iron abundances ([Fe/H]), age ($t$), proper motion components ($\langle\mu_{\alpha}\cos\delta\rangle$, $\langle\mu_{\delta}\rangle$), and radial velocity ($V_{\gamma}$).}
  \begin{tabular}{ccccccccc}
    \hline
    \hline
    \multicolumn{9}{c}{Berkeley 68}\\
        \hline
        \hline
$E(B-V)$ & $\mu$ & $d$ & [Fe/H] & $t$ &  $\langle\mu_{\alpha}\cos\delta\rangle$ &  $\langle\mu_{\delta}\rangle$ & $V_{\rm R}$ & Ref \\
(mag) & (mag) & (pc)  & (dex) & (Myr) & (mas yr$^{-1}$) & (mas yr$^{-1}$) & (km s$^{-1})$ &      \\
    \hline
 0.67 & 11.492 & 1800                  & ---    & 1300        & -5.860 & +0.300  & --- & (1) \\
 ---  & ---    & $3223_{-786}^{+1531}$ & ---    &  ---        & +2.310$\pm$0.012 & -1.312$\pm$0.008 & --- & (2) \\
 ---  & ---    & 3223                  & ---    & ---         & +2.310$\pm$0.012 & -1.312$\pm$0.008 & -20.31$\pm1.86$ & (3)\\
 ---  & ---    & 3559$\pm$608          & ---    & 1450$\pm$90 & +2.321$\pm$0.227 & -1.322$\pm$0.203 & --- & (4) \\  
0.445 & 12.65  & 3383                  & ---    & 2100        & +2.310$\pm$0.012 & -1.312$\pm$0.008 & --- & (5) \\
---   & ---    & 3223                  & -0.183$\pm$0.168     & ---              & +2.310$\pm$0.166 & -1.312$\pm$0.135 & -29.09$\pm9.30$ & (6)\\

0.52$\pm$0.04  & 13.7$\pm$0.2          & 2554$\pm$387         & ---              & 1800$\pm$200     & +2.29$\pm$0.37   & -1.29$\pm$0.24  & --- & (7) \\
---   & ---    & 3261                  & ---    & 2140        & +2.310$\pm$0.012 & -1.312$\pm$0.008 & -63.64$\pm16.23$& (8)\\
---   & ---    & 3356$\pm$225          & ---    & 1300        & +2.188$\pm$0.056 & -1.353$\pm$0.048 & -18.61$\pm0.13$& (9)\\
---   & 12.65  & 3425$\pm$540        & ---    & 2100        & +2.242$\pm$0.079 & -1.393$\pm$0.053 & --- & (10) \\
0.67$\pm$0.06 &---&3027$\pm$118        & -0.286$\pm$0.155& 1900$\pm$370 & +2.357$\pm$0.194 & -1.290$\pm$0.133 & -21.54$\pm$0.99& (11) \\
0.520$\pm$0.032& 14.00$\pm$0.12 & 3003$\pm$165 & -0.13$\pm$0.08& 2400$\pm$200 & +2.237$\pm$0.007  & -1.401$\pm$0.005  & -20.31$\pm$1.86 & (12) \\
  \hline
  \hline
    \multicolumn{9}{c}{Stock 20}\\
        \hline
        \hline
$E(B-V)$ & $\mu$ & $d$ & [Fe/H] & $t$ &  $\langle\mu_{\alpha}\cos\delta\rangle$ &  $\langle\mu_{\delta}\rangle$ & $V_{\rm R}$ & Ref \\
    (mag) & (mag) & (pc)  & (dex) & (Myr) & (mas yr$^{-1}$) & (mas yr$^{-1}$) & (km s$^{-1}$) & \\
    \hline
 0.20 & 10.271 & 1100 & --- &  220 & -2.560 & -3.240 & --- & (1) \\
 ---  & ---    & $2692_{-571}^{+990}$ & --- &  --- & -3.237$\pm$0.010 & -1.124$\pm$0.011 & --- & (2) \\
  ---  & ---    & 2890$\pm$309 & --- & 8$\pm$1 & -3.164$\pm$0.258 & -0.950$\pm$0.308 & --- & (4) \\  
0.352 & 12.07  & 2598 & --- &  19 & -3.237$\pm$0.010 & -1.124$\pm$0.011 & --- & (5) \\
---   & ---    & 2674$\pm$157          & ---    & 220        & -2.523$\pm$0.162 & -0.932$\pm$0.078 & --- & (9)\\
--- & 12.07  & 2786$\pm$109 & --- &  19 & -3.209$\pm$0.042 & -1.217$\pm$0.066 & --- & (10) \\
0.48$\pm$0.01 &---&2566$\pm$230        & -0.100$\pm$0.117& 13$\pm$3 & -3.235$\pm$0.072 & -1.129$\pm$0.083 & --- & (11) \\
0.400$\pm$0.048& 13.56$\pm$0.16 & 2911$\pm$216 & -0.01$\pm$ 0.06 & 50$\pm$10 & -3.215$\pm$0.004 & -1.172$\pm$0.004  & --- & (12) \\
  \hline
    \end{tabular}%
    \\
(1) \citet{Kharchenko12}, (2) \citet{Cantat-Gaudin18}, (3) \citet{Soubiran18}, (4) \citet{Liu19}, (5) \citet{Cantat-Gaudin20}, (6) \citet{Zhong20}, 
(7) \citet{Maurya20}, (8) \citet{Tarricq21}, (9) \citet{Hao21}, (10) \citet{Poggio21}, (11) \citet{Dias21}, (12) This study
  \label{tab:literature}%
\end{table*}%

% FIGURE 1
\begin{figure*}
\centering
\includegraphics[scale=0.18, angle=0]{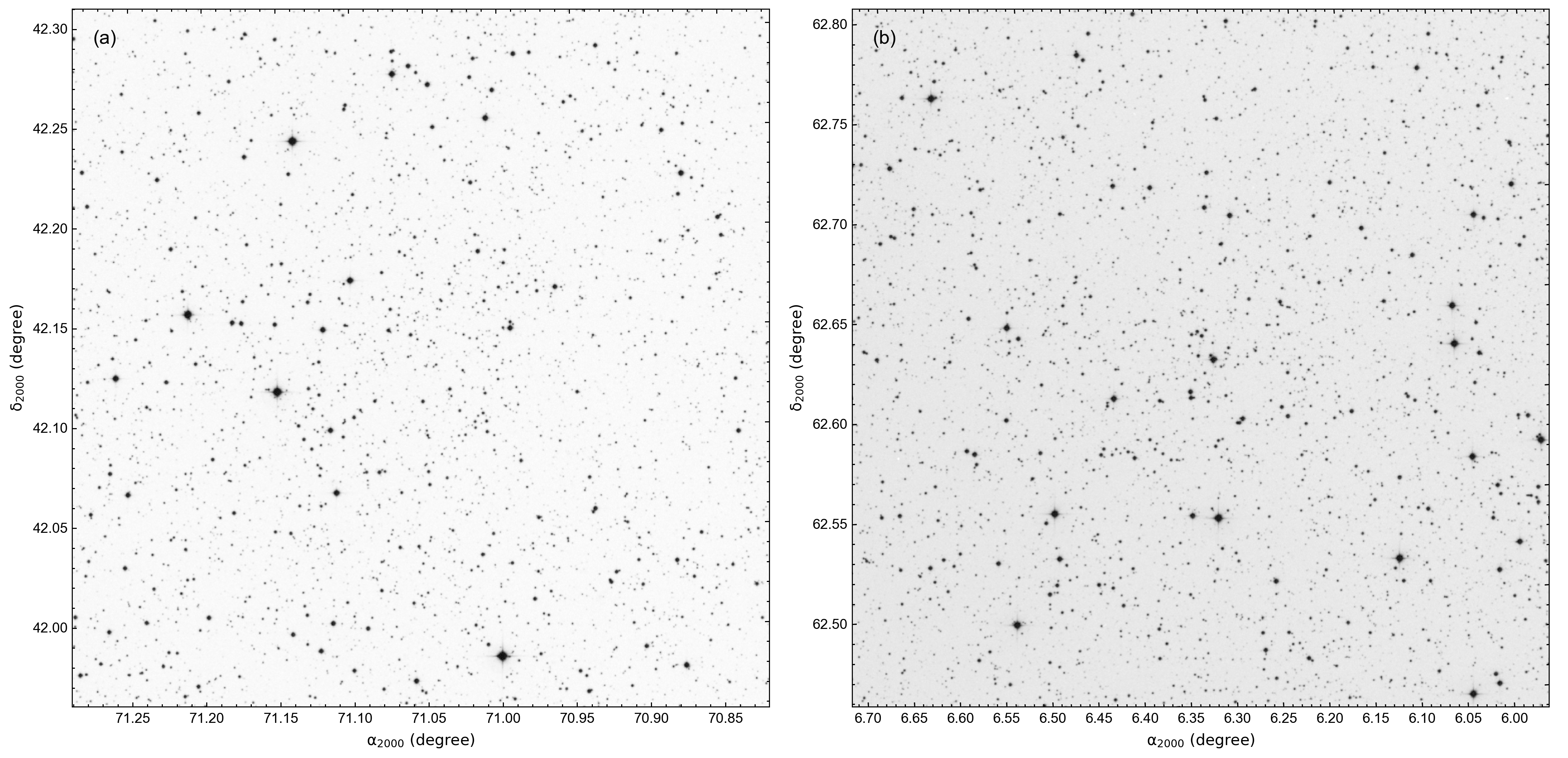}
\caption{Identification chart of stars located in regions of Berkeley 68 (a) and for Stock 20 (b). Field of view of the charts is $21'.5 \times 21'.5$. North and East correspond
to the up and left directions, respectively} 
\label{fig:charts}
\end {figure*}

%---------------------------------------------------------------

% Table 2
\begin{table}[t]
\setlength{\tabcolsep}{4.5pt}
\renewcommand{\arraystretch}{0.6}
  \centering
  \caption{Observation log for Berkeley 68 and Stock 10. Columns give the cluster names, date of observation (day-month-year), and photometric filters. Under `filters' rows give the exposure times (in seconds) and the number of exposures ($N$).}
  \medskip
    \begin{tabular}{cclll}
    \hline
            &  & \multicolumn{3}{c}{Filter/Exp, Time (s) $\times N$}   \\
    \hline
    Cluster & Obs. Date  & $U$           & $B$          & $V$          \\
    \hline
Berkeley 68 & 06-11-2018 &   60$\times$3 &   6$\times$6 &   3$\times$5 \\
            &            & 1800$\times$2 & 600$\times$3 & 360$\times$3 \\
Stock 20    & 08-10-2016 &   90$\times$2 &   8$\times$4 &   4$\times$5 \\
            &            & 1800$\times$2 & 600$\times$2 & 300$\times$3 \\
    \hline
    \end{tabular}%
  \label{tab:exposures}%
\end{table}%

%-------------------------------------------------------------------------------------------
\section{Observations}

CCD {\it UBV} observations of stars located through the cluster regions of Berkeley 68 were made on 2018 November 06 and on 2016 October 08 for Stock 20. We used the T100 100-cm $f/10$ Ritchey-Chr\'etien telescope at the T\"UB\.ITAK National Observatory (TUG)\footnote{www.tug.tubitak.gov.tr} in Turkey. Images were taken with a back illuminated 4k$\times$4k pixel CCD. The image scale is $0''\!\!.31$ pixel$^{-1}$ with a total field of view of $21'\!\!.5 \times 21'\!\!.5$. The CCD gain was 0.55 e$^{-}$/ADU, while readout noise was 4.19~e$^{-}$ (100 KHz). Identification charts are shown in Figure~\ref{fig:charts} and a log of observations is listed in Table~\ref{tab:exposures}.  The shorter exposures were to obtain unsaturated images of the brighter stars, while the longer exposures were aimed at collecting good photometry of the fainter stars.  Standard stars of \citet{Landolt09} were observed on the same nights as the two clusters and used for photometric calibration. 16 Landolt regions were observed over an airmass range between 1.23 and 1.90, providing a total of 111 standard stars (see Table~\ref{tab:standard_stars}). Standard bias subtraction and flat fielding procedures inside IRAF\footnote{IRAF is distributed by the National Optical Astronomy Observatories} were applied to all the raw science images. The instrumental magnitudes of the \citet{Landolt09} stars were calculated using IRAF's aperture photometry packages. Applying multiple linear regression fits to these magnitudes led to estimates of the photometric extinction and transformation coefficients for the two observing nights. Derived values were listed in Table~\ref{tab:coefficients}. For the cluster images, we performed astrometric corrections via PyRAF\footnote{PyRAF is a product of the Space Telescope Science Institute, which is operated by AURA for NASA} and astrometry.net\footnote{http://astrometry.net}. Then we measured the instrumental magnitudes of the objects detected in cluster areas using Source Extractor (SExTractor) and PSF Extractor (PSFEx) routines \citep{Bertin96}. Aperture corrections were applied to these magnitudes. After this step, we transformed instrumental magnitudes to standard $U\!BV$ magnitudes considering transformation equations of \citet{Janes11}:   

% Table 3
\begin{table}[t]
\setlength{\tabcolsep}{3pt}
\renewcommand{\arraystretch}{0.6}
  \centering
  \caption{Selected \citet{Landolt09} standard star fields. The columns denote the observation date (day-month-year), star field name from Landolt, the number of standard stars ($N_{\rm st}$) observed in a given field, the number of observations for each field ($N_{\rm obs}$), and the airmass range the fields were observed over ($X$).}
\medskip
    \begin{tabular}{llccc}
    \hline
Date	   & Star Field	& $N_{\rm st}$ & $N_{\rm obs}$	& $X$          \\
\hline
      	   & SA92SF3    &  6 	       & 2	        &              \\
      	   & SA93       &  4	       & 1	        &              \\
      	   & SA95SF2	&  9	       & 1	        &              \\
      	   & SA97SF1    &  8	       & 1	        &              \\
08-10-2016 & SA98	    & 19	       & 1	        & 1.24 -- 1.90 \\
           & SA111	    &  5	       & 1	        &              \\
           & SA112	    &  6	       & 2	        &              \\
	       & SA113	    & 15	       & 2	        &              \\
	       & SA114	    &  5	       & 2	        &              \\
\hline	      
           & SA92       &  6           & 1	        &              \\
	       & SA93       &  4	       & 1	        &              \\
	       & SA94	    &  2 	       & 1	        &              \\
           & SA95       &  9	       & 1	        &              \\
06-11-2018 & SA96	    &  2	       & 2	        &              \\
	       & SA97       &  2	       & 2	        &              \\
	       & SA98	    & 19	       & 1	        & 1.23 -- 1.87 \\
	       & SA99       &  3	       & 1	        &              \\
	       & SA110	    & 10	       & 1	        &              \\
	       & SA111	    &  5	       & 1	        &              \\	
	       & SA112	    &  6	       & 1	        &              \\
	       & SA114	    &  5	       & 1	        &              \\	       
    \hline
    \end{tabular}%
  \label{tab:standard_stars}%
\end{table}%

% Table 4
\begin{table*}
\renewcommand{\arraystretch}{0.6}
  \centering
  \caption{Transformation and extinction coefficients obtained for the two observation nights: $k$ and $k'$ are the primary and secondary extinction coefficients, while $\alpha$ and $C$ are the transformation coefficients. Dates are day-month-year.} 
  \medskip
    \begin{tabular}{lccccc}
    \hline
Filter/color index & Obs. Date & $k$   & $k'$             & $\alpha$          & $C$             \\
    \hline
$U$     & 08.10.2016 & 0.308$\pm$0.061 & +0.091$\pm$0.110 & ---               & ---             \\
$B$     &            & 0.151$\pm$0.041 & -0.003$\pm$0.048 & 0.900$\pm$0.073   & 1.444$\pm$0.063 \\
$V$     &            & 0.104$\pm$0.018 & ---              & ---               & ---             \\
$U-B$   &            & ---             & ---              & 0.754$\pm$0.158   & 3.807$\pm$0.089 \\
$B-V$   &            & ---             & ---              & 0.068$\pm$0.009   & 1.391$\pm$0.027 \\
$U$     & 06.11.2018 & 0.521$\pm$0.062 & -0.122$\pm$0.069 & ---               & ---             \\ 
$B$     &            & 0.227$\pm$0.046 & -0.021$\pm$0.052 & 0.932$\pm$0.077   & 1.512$\pm$0.068 \\
$V$     &            & 0.124$\pm$0.018 & ---              & ---               & ---             \\
$U-B$   &            & ---             & ---              & 0.987$\pm$0.102   & 3.718$\pm$0.093 \\
$B-V$   &            & ---             & ---              & 0.073$\pm$0.007   & 1.570$\pm$0.028 \\
\hline
    \end{tabular}%
  \label{tab:coefficients}%
\end{table*}%

\begin{equation}
V = v - \alpha_{bv}(B-V)-k_vX _v- C_{bv} \\
\end{equation}

\begin{equation}
B-V = \frac{(b-v)-(k_b-k_v)X_{bv}-(C_b-C_{bv})}{\alpha_b+k'_bX_b-\alpha_{bv}} \\
\end{equation}

\begin{eqnarray}
U-B = \frac{(u-b)-(1-\alpha_b-k'_bX_b)(B-V)}{\alpha_{ub}+k'_uX_u}-\frac{(k_u-k_b)X_{ub}-(C_{ub}-C_b)}{\alpha_{ub}+k'_uX_u} 
\end{eqnarray}
In equations (1-3) $U\!$, $B$, and $V$ are the magnitudes in the standard photometric system. $u$, $b$, and $v$ indicate the instrumental magnitudes. $X$ is the airmass. $k$ and $k'$ represent primary and secondary extinction coefficients. $\alpha$ and $C$ are transformation coefficients to the standard system.

%---------------------------------------------------------------
%\clearpage
\section{Data Analysis}
\subsection{Photometric data}

We constructed {\it UBV} photometric, {\it Gaia} DR3 photometric, and astrometric catalogues for the detected stars in the direction of Berkeley 68 (3,548 stars) and Stock 20 (2,866 stars). Both of catalogues are available electronically\footnote{The complete tables can be obtained from VizieR electronically} and contain the following information for the detected stars: IDs, positions ($\alpha, \delta$), $UBV$-based $V$ apparent magnitudes and color indices ($U-B$, $B-V$), {\it Gaia} DR3-based apparent $G$ magnitudes and color index ($G_{\rm BP}-G_{\rm RP}$), proper motion components ($\mu_{\alpha}\cos\delta, \mu_{\delta}$) along with trigonometric parallaxes ($\varpi$), and membership probabilities ($P$) as calculated in this study (Table~\ref{tab:all_cat}). Inaccuracies for  the $V$, $U-B$, $B-V$, $G$, $G_{\rm BP}-G_{\rm RP}$ magnitudes and color indices were adopted as internal errors. The mean internal photometric errors per $V$ magnitude bin in the Johnson and {\it Gaia} DR3 filters are listed in Table~\ref{tab:phiotometric_errors} (on page~\pageref{tab:phiotometric_errors}). This table shows that the {\it UBV} errors are smaller than 0.2 mag for Berkeley 68, while the errors reach up 0.4 mag in Stock 20 for stars fainter than $V=23$ mag. The {\it Gaia} DR3 photometric mean errors are smaller than 0.2 mag for stars down to $V=23$ mag for both clusters.  

\begin{figure*}
\centering
\includegraphics[scale=0.4, angle=0]{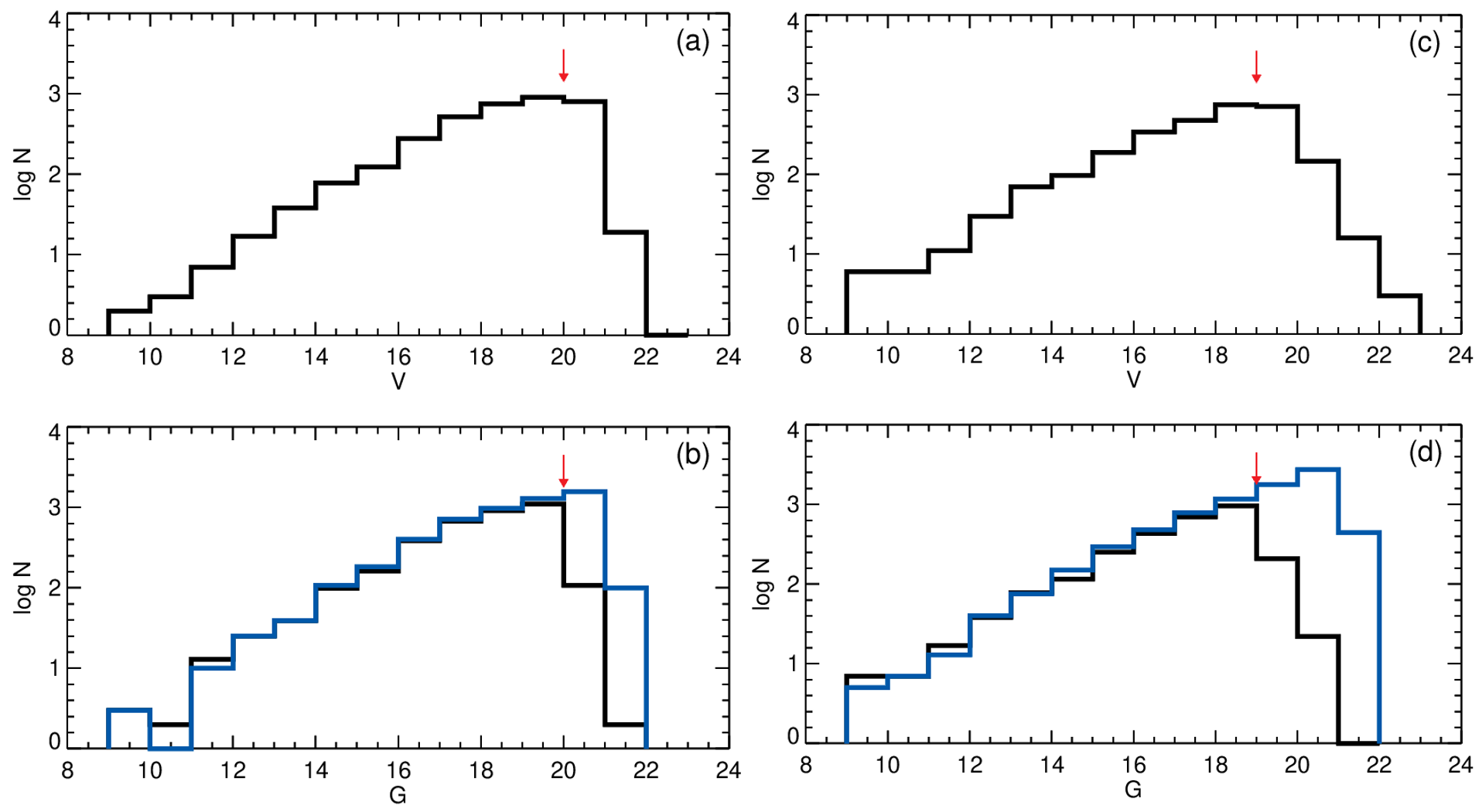}\\
\caption{Histograms of Berkeley 68 (a, b) and Stock 20 (c, d) for per magnitude bin in $V$ and $G$ filters. The red arrows indicate the adopted faint limiting apparent magnitudes in $V$ and $G$ filters. Black lines are the star counts based on $V$ band photometry, while blue lines are counts based on {\it Gaia} DR3 data for the same cluster regions.} 
\label{fig:histograms}
\end {figure*}

Increasingly faint stars are not always able to be reliably detected in CCD frames. This is described as decreasing completeness of the stellar counts as magnitude increases.  The photometric completeness limit is necessary for further analyses investigating a cluster's luminosity function, mass function, stellar density distribution, etc. To investigate photometric completeness for both clusters, we constructed $V$ and $G$ magnitude histograms which we then compared with those based on the {\it Gaia} DR3 data for the same regions on the sky as the CCD images. In preparing the {\it Gaia} DR3 data, we considered central equatorial coordinates presented by \citet{Cantat-Gaudin20} and limited stellar magnitudes to the range $8<G<23$ mag. Distributions of all the stellar counts are shown in Fig.~\ref{fig:histograms}: the black solid lines indicate the observational stellar distributions by magnitude bin in $V$ and $G$, while the blue solid lines (see in Fig.~\ref{fig:histograms}b and \ref{fig:histograms}d) show the counts based on {\it Gaia} DR3. Figures~\ref{fig:histograms}b and \ref{fig:histograms}d demonstrate that the number of stars detected in the study are well aligned with the stellar counts from {\it Gaia} DR3 data up to fainter magnitudes adopted as being the completeness limits (the red vertical arrows in Figures~\ref{fig:histograms}b and \ref{fig:histograms}d). These limits are $V=20$ mag for Berkeley 68 and $V=19$ mag for Stock 20, which corresponds to $G=20$ mag and $G=19$ mag, respectively. From Figures~\ref{fig:histograms}b and \ref{fig:histograms}d, one can see that for magnitudes fainter than the adopted completeness limits, stellar counts decrease the due to the increased crowding of low-mass stars. Different telescope-detector combinations and telescope properties of the ground and space based observations affect the detected stars particularly at fainter magnitudes. This could explain the greater number of detected stars fainter than $G>19$ mag in the {\it Gaia} space-based observations.    

% FIGURE 2

% TABLE 5
% otherwise.  
\begin{sidewaystable}
\setlength{\tabcolsep}{1.9pt}
\renewcommand{\arraystretch}{0.8}
\footnotesize
  \centering
 \caption{The catalogues for Berkeley 68 and Stock 20. The complete table can be found electronically.}
    \begin{tabular}{cccccccccccc}
\hline
\multicolumn{11}{c}{Berkeley 68}\\
\hline
ID	 & RA           &	DEC	        &      $V$	    &	$U-B$      & $B-V$	      &	$G$	          & 	$G_{\rm BP}-G_{\rm RP}$	 & 	$\mu_{\alpha}\cos\delta$ & 	$\mu_{\delta}$ & 	$\varpi$	& $P$ \\

	 & (hh:mm:ss.ss)           &	(dd:mm:ss.ss)	&      (mag)	    &	(mag)      & (mag)	      &	(mag)	          & 	(mag)	 & 	(mas yr$^{-1}$) & 	(mas yr$^{-1}$) & 	(mas)	&  \\
\hline
0001 & 04:43:31.37 & +42:10:45.82 & 19.531(0.022) & ---          & 1.013(0.031) & 19.078(0.004) & 1.370(0.067) & 0.895(0.451) & -0.600(0.294) & 0.561(0.343) & 0.00 \\
0002 & 04:43:31.48 & +42:07:57.25 & 20.536(0.054) & ---          & 1.526(0.097) & 19.381(0.004) & 2.048(0.088) &-1.624(0.617) & -5.824(0.391) & 1.114(0.472) & 0.00 \\
0003 & 04:43:31.50 & +42:00:22.73 & 19.561(0.024) & ---          & 1.072(0.035) & 18.880(0.004) & 1.642(0.047) &-0.140(0.328) &  1.851(0.227) & 0.162(0.255) & 0.00 \\
0004 & 04:43:31.50 & +42:12:49.47 & 18.189(0.008) & 0.572(0.041) & 1.108(0.012) & 17.887(0.003) & 1.465(0.025) & 0.880(0.163) &  0.414(0.124) & 0.523(0.118) & 0.04 \\
...  & ...         & ...	  & ...           & ...          & ...          & ...           & ...          & ...          & ...           & ...          & ... \\
...  & ...         & ...	  & ...           & ...          & ...          & ...           & ...          & ...          & ...           & ...          & ... \\
...  & ...         & ...	  & ...           & ...          & ...          & ...           & ...          & ...          & ...           & ...          & ... \\
3545 &04:45:23.12  &+42:08:49.52   &19.181(0.016)  & ---          &1.181(0.024)  &18.558(0.003)  &1.658(0.042)  &1.117(0.208)  &-1.758(0.173)  &0.257(0.177)  & 0.00\\
3546 &04:45:23.17   &+42:04:39.92   &19.362(0.018)  & ---          &1.442(0.030)  &18.718(0.003   &1.732(0.048)  &0.734(0.313)  &-3.046(0.222)  &0.720(0.222)  & 0.00\\
3547 &04:45:23.27  &+42:03:11.99   &20.438(0.046)  & ---          &1.267(0.071)  &19.767(0.005)  &1.543(0.080)  &1.602(0.553)  &-1.178(0.451)  &0.337(0.476)  & 0.00 \\
3548 &04:45:23.32  &+42:11:02.03    &19.068(0.014)  &0.591(0.081)  &1.092(0.021)  &18.651(0.003)  &1.475(0.040)  &0.701(0.212)  &-1.410(0.184)  &0.537(0.185)  & 0.00 \\
\hline
\multicolumn{11}{c}{Stock 20}\\
\hline
ID	 & RA           &	DEC	        &      $V$	    &	$U-B$      & $B-V$	      &	$G$	          & 	$G_{\rm BP}-G_{\rm RP}$	 & 	$\mu_{\alpha}\cos\delta$ & 	$\mu_{\delta}$ & 	$\varpi$	& $P$ \\
	 & (hh:mm:ss.ss)           &	(dd:mm:ss.ss)	&      (mag)	    &	(mag)      & (mag)	      &	(mag)	          & 	(mag)	 & 	(mas yr$^{-1}$) & 	(mas yr$^{-1}$) & 	(mas)	&  \\
\hline
0001 & 00:23:44.36  & +62:39:25.57 & 18.724(0.015)  & 0.666(0.069) & 1.185(0.025)  & 18.086(0.003) & 1.768(0.019) &-1.646(0.090) & -0.562(0.099) &  0.237(0.099)  & 0.01 \\
0002 & 00:23:44.65  & +62:42:56.70 & 17.956(0.008)  & 0.796(0.041) & 1.212(0.014)  & 17.213(0.003) &1.956(0.017)  &-1.380(0.060) & -0.673(0.068) &  0.233(0.061)  & 0.00 \\
0003 & 00:23:44.71  & +62:44:58.57 & 15.917(0.002)  & 0.396(0.006) & 0.914(0.003)  & 16.340(0.003) &1.258(0.035)  &-5.383(0.080) & -3.671(0.080) &  0.603(0.080)  & 0.00 \\
0004 & 00:23:45.14  & +62:28:14.77 & 22.509(0.436)  & ---          & 0.122(0.505)  & 20.623(0.008) &2.485(0.124)  & 3.072(0.882) & -2.728(0.802) &  0.108(0.789)  & 0.00 \\
...  & ...          & ...	   & ...            & ...          & ...           & ...           & ...          & ...          & ...           & ...            & ...  \\
...  & ...          & ...	   & ...            & ...          & ...           & ...           & ...          & ...          & ...           & ...            & ...  \\
...  & ...          & ...	   & ...            & ...          & ...           & ...           & ...          & ...          & ...           & ...            & ...  \\
2863 & 00:26:43.22  & +62:44:26.22 & 20.068(0.047)  & ---          & 0.601(0.061)  & 18.890(0.003) & 1.750(0.068) & -1.349(0.151)& -0.258(0.178) & -0.064(0.172)  & 0.00 \\
2864 & 00:26:43.34  & +62:44:48.44 & 18.878(0.017)  & 0.596(0.073) & 1.158(0.028)  & 18.281(0.003) & 1.593(0.033) & -2.915(0.101)&  0.304(0.112) &  0.223(0.112)  & 0.00 \\
2865 & 00:26:43.57  & +62:38:22.89 & 19.851(0.038)  & ---          & 1.335(0.068)  & 19.154(0.004) & 1.728(0.053) & -2.490(0.173)& -1.053(0.211) &  0.087(0.191)  & 0.01 \\
2866 & 00:26:43.67  & +62:46:03.67 & 20.590(0.088)  & ---          & 1.710(0.180)  & 19.774(0.004) & 2.155(0.098) &  0.796(0.296)&  0.725(0.308) & -0.105(0.308)  & 0.00 \\
\hline
    \end{tabular}
      \label{tab:all_cat}%
\end{sidewaystable} 

% Table 6
\begin{table*}
\setlength{\tabcolsep}{3pt}
\renewcommand{\arraystretch}{0.8}
  \centering
  \caption{Mean internal photometric errors per magnitude bin in $V$ brightness.}
    \begin{tabular}{ccccccc|ccccccc}
      \hline
    \multicolumn{7}{c}{Berkeley 68} & \multicolumn{6}{c}{Stock 20} \\
    \hline
  $V$ & $N$ & $\sigma_{\rm V}$ & $\sigma_{\rm U-B}$ & $\sigma_{\rm B-V}$ & $\sigma_{\rm G}$ &  $\sigma_{G_{\rm BP}-G_{\rm RP}}$ & $V$ & $N$ & $\sigma_{\rm V}$ & $\sigma_{\rm U-B}$ & $\sigma_{\rm B-V}$ & $\sigma_{\rm G}$ & $\sigma_{G_{\rm BP}-G_{\rm RP}}$\\
  \hline
  (08, 12] &  12 & 0.002 & 0.002 & 0.002 & 0.004 & 0.005 & (08, 12] &  23 & 0.001 & 0.002 & 0.002 & 0.003 & 0.005\\
  (12, 13] &  17 & 0.003 & 0.004 & 0.004 & 0.003 & 0.005 & (12, 13] &  30 & 0.002 & 0.003 & 0.003 & 0.003 & 0.005\\
  (13, 14] &  38 & 0.005 & 0.005 & 0.006 & 0.003 & 0.006 & (13, 14] &  70 & 0.004 & 0.003 & 0.005 & 0.003 & 0.005\\
  (14, 15] &  78 & 0.002 & 0.005 & 0.002 & 0.003 & 0.005 & (14, 15] &  98 & 0.001 & 0.003 & 0.001 & 0.003 & 0.006\\
  (15, 16] & 124 & 0.002 & 0.006 & 0.002 & 0.003 & 0.006 & (15, 16] & 189 & 0.002 & 0.006 & 0.003 & 0.003 & 0.006\\
  (16, 17] & 277 & 0.002 & 0.010 & 0.004 & 0.003 & 0.007 & (16, 17] & 342 & 0.003 & 0.011 & 0.005 & 0.003 & 0.007\\
  (17, 18] & 518 & 0.005 & 0.021 & 0.007 & 0.003 & 0.012 & (17, 18] & 481 & 0.006 & 0.025 & 0.010 & 0.003 & 0.011\\
  (18, 19] & 753 & 0.009 & 0.045 & 0.014 & 0.003 & 0.026 & (18, 19] & 757 & 0.012 & 0.054 & 0.022 & 0.003 & 0.020\\
  (19, 20] & 909 & 0.020 & 0.068 & 0.033 & 0.004 & 0.054 & (19, 20] & 711 & 0.028 & 0.080 & 0.054 & 0.003 & 0.037\\
  (20, 21] & 802 & 0.042 & 0.080 & 0.071 & 0.005 & 0.099 & (20, 21] & 146 & 0.055 & 0.395 & 0.110 & 0.004 & 0.058\\
  (21, 23] &  20 & 0.090 &  ---  & 0.168 & 0.008 & 0.181 & (21, 23] &  19 & 0.201 &  ---  & 0.353 & 0.012 & 0.189\\
      \hline
    \end{tabular}%
  \label{tab:phiotometric_errors}%
\end{table*}%

\subsection{The Radial Density Profile}

\begin{figure}[t]
\centering
\includegraphics[scale=0.5, angle=0]{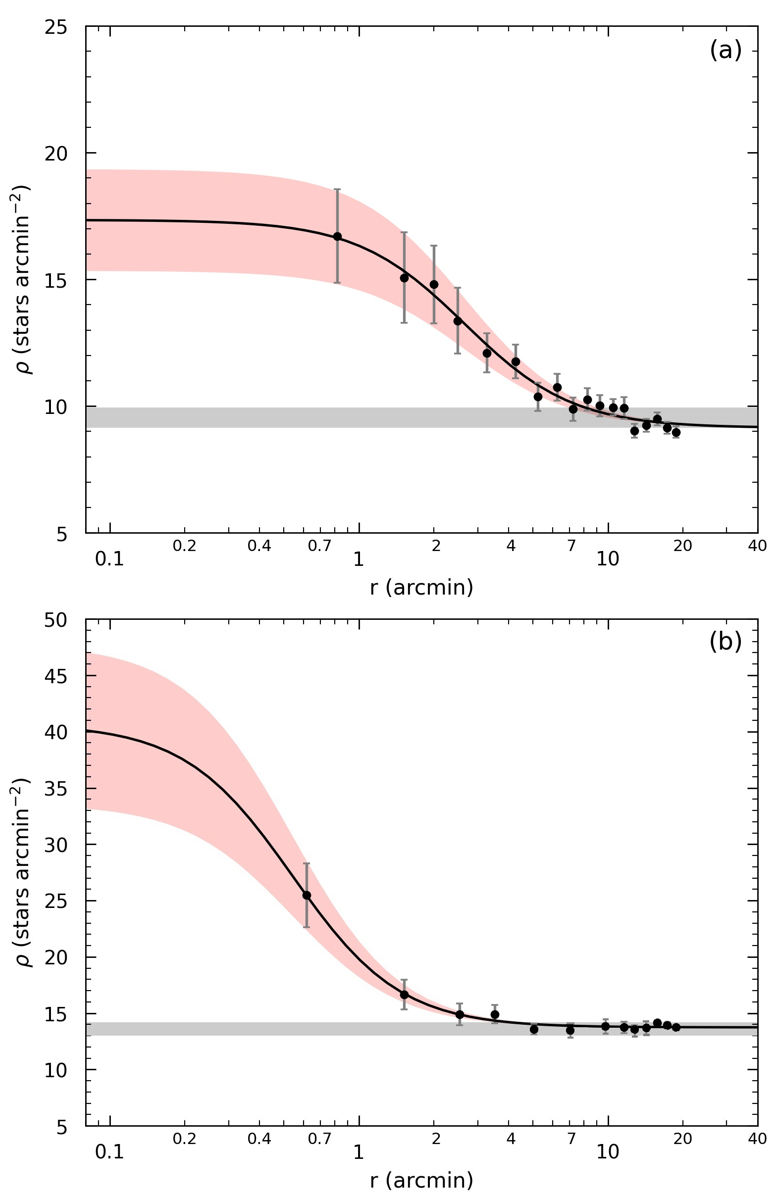}\\
\caption{The stellar density distribution for Berkeley 68 (a) and Stock 20 (b). The fitted black curve represents the \citet{King62} profile, while the horizontal grey band shows background stellar density. The $1\sigma$ King fit uncertainty is shown by the red shaded domain.} 
\label{fig:king}
\end {figure}

Varying stellar distributions at different magnitudes can affect the spatial analysis of a cluster. We adopted central coordinates as given by \citet{Cantat-Gaudin20} for both clusters. To construct a radial density profile (RDP) we considered the stars brighter than the photometric completeness limits ($V=20$ mag for Berkeley 68 and $V=19$ for Stock 20) and estimated the stellar density in a series of concentric rings centered on the cluster centers within equal incremental steps of 1 arcmin. To calculate stellar densities ($\rho$), we divided star counts in each ring by that ring's area. The stellar density uncertainties were derived considering Poisson statistics ($1/\sqrt N$, where $N$ represents the number of stars used in the density estimation). We applied the empirical RDP of \citet{King62} $\rho(r)=f_{\rm bg}+ [f_{\rm 0}/(1+(r/r_{\rm c})^2)]$. Variables $r$, $f_{\rm 0}$,  $r_{\rm c}$, and  $f_{\rm bg}$ are the radius from the cluster centre, the central density, core radius, and background density. The fitting procedure utilized $\chi^2$ minimization. Figure~\ref{fig:king} shows the stellar density distribution by radius from the cluster centre  based on the best fit model to the RDP. The values of the central stellar density and core radius of the clusters, together with the background stellar density, were found to be $f_{\rm 0}=8.204\pm 2.008$ stars arcmin$^{-2}$, $r_{\rm c}=2.663\pm 0.616$ arcmin, and $f_{\rm bg}=9.139\pm 0.402$ stars arcmin$^{-2}$ for Berkeley 68 and $f_{\rm 0}=27.197\pm 7.164$ stars arcmin$^{-2}$, $r_{\rm c}=0.543\pm 0.225$ arcmin, and $f_{\rm bg}=13.742\pm 0.222$ stars arcmin$^{-2}$ for Stock 20. It can be seen in Fig.~\ref{fig:king} that the background density (shown by the horizontal grey  band) crosses the RDP model at the limiting radius for both clusters. These limiting radii are adopted as $r_{\rm lim}=8'$ ($6.99 \pm 0.38$ pc) for Berkeley 68 and $r_{\rm lim}=7'.5$ ($6.35 \pm 0.47$ pc) for Stock 20. For subsequent analyses, we considered only the stars inside these limiting radii.

% FIGURE 3

%----------------------------------------------------------------------------------
\clearpage
\subsection{CMDs and Cluster Membership}
\label{section:cmds}

It is important to remove the field star contamination so that subsequent analysis is on likely cluster member stars only. Due to their formation from the same source, open cluster stars share common vectoral movements in the sky. For this reason, study of stellar proper motion components is an important tool to assess cluster membership for these stars \citep{Yadav13,Bisht20}. In order to specify membership probabilities of stars, various researchers have set up mathematical models and statistical methods that depend on analyses of stellar proper motion components \citep{Vasilevskis58, Stetson80, Zhao90, Balaguer98}.

In the study, we applied the Unsupervised Photometric Membership Assignment in Stellar Clusters \citep[{\sc UPMASK};][]{Krone-Martins14} method and used {\it Gaia} DR3 astrometric data to calculate membership probabilities ($P$) of stars in the regions of Berkeley 68 and Stock 20. This method was successfully applied in various earlier studies to identify open cluster members \citep{Cantat-Gaudin18, Cantat-Gaudin20, Castro-Ginard18, Castro-Ginard19, Castro-Ginard20, Banks20, Akbulut21, Wang22}. {\sc UPMASK} is based on the $k$-means clustering method  (where $k$ is the number of data clusters) which considers an open stellar cluster as a group of stars which share a similar space in proper motion components and parallaxes. The $k$-means algorithm allows detection of spatially concentrated groups, and hence membership groupings of stars are identified. Best results are typically reached when the $k$-means value is within the range 6 to 25 inclusive. In order to identify likely members of the Berkeley and Stock 20, we utilized {\sc UPMASK} taking into account five-dimensional astrometric data ($\alpha$, $\delta$, $\mu_{\alpha}\cos \delta$, $\mu_{\delta}$, $\varpi$) and associated uncertainties for each detected star. We ran 100 iterations for both of the clusters to evaluate membership probabilities.  We obtained 298 likely member stars for Berkeley 68 and 107 for Stock 20 when we applied the membership probability $P\geq0.5$ and the completeness $V$ magnitude limit derived above for both clusters (see Section 3.)

\begin{figure*}[t]
\centering
\includegraphics[scale=0.70, angle=0]{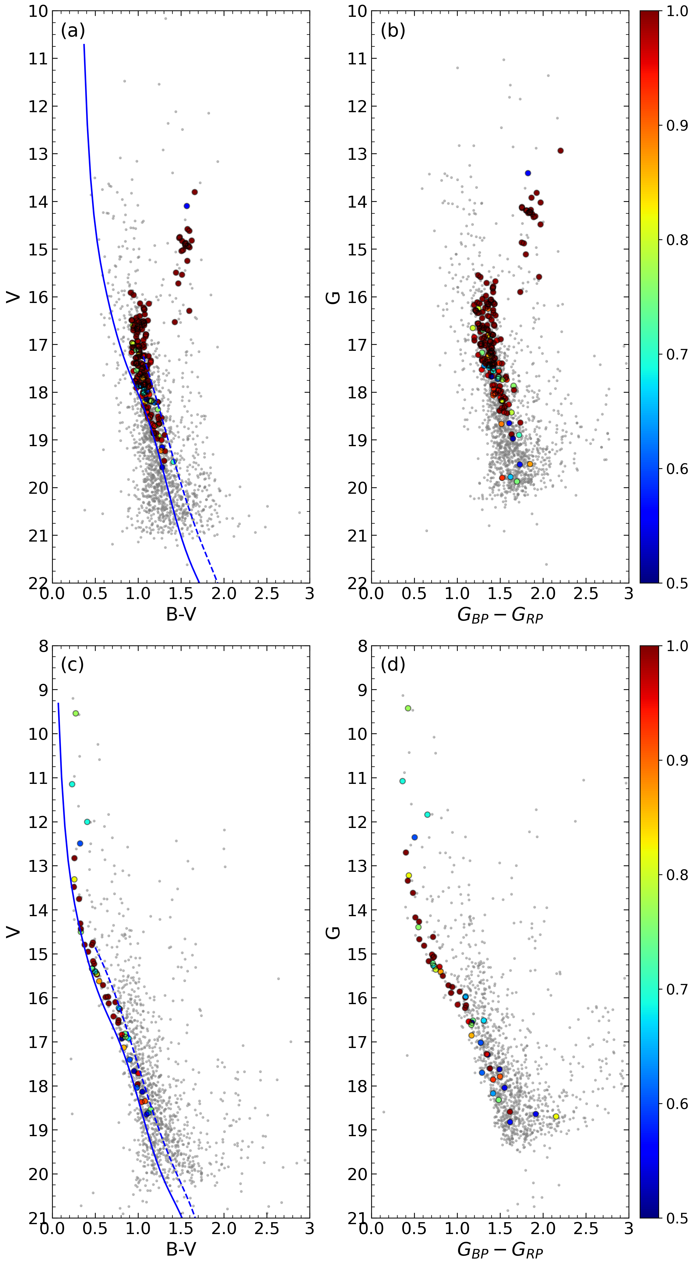}
\caption{CMDs of Berkeley 68 (a, b) and Stock 20 (c, d) based on {\it UBV} (a, c) and {\it Gaia} DR3 photometry (b, d). The blue dashed lines show the ZAMS \citep{Sung13} including the binary star contamination. The membership probabilities of stars are shown with different colors, these member stars are located within $r_{\rm lim}=8'$ and $r_{\rm lim}=7'\!.5$ of the cluster centres calculated for Berkeley 68 and Stock 20, respectively. The stars with the low membership probabilities were indicated with grey dots. 
\label{fig:cmds}}
\end {figure*}

The CMDs of stellar clusters are effective tools for estimating fundamental parameters such as age, distance, reddening etc. They are also helpful for the separation of cluster members from field stars, along with exploring binary-star contamination in cluster main-sequences. Therefore, to eliminate possible `contamination' of the observed main sequence by binary stars in Berkeley 68 and Stock 20, we constructed $V\times (B-V)$ CMDs from the stars located inside the limiting radii ($r_{\rm lim}$). Marking most probable cluster members ($P\geq0.5$) on the diagrams we fitted a Zero Age Main-Sequence (ZAMS) curve based on \citet{Sung13} to the CMDs by eye. Reference was made to probable members that are located at the fainter end of the observed cluster main-sequences. To account for main-sequence binary star contamination, we shifted the ZAMS by 0.75 mag towards brighter magnitudes to ensure that we selected the most probable main-sequence, turn-off and giant members of each cluster. Hence, considering the completeness-derived $V$ magnitude limits, cluster $r_{\rm lim}$ radii, and ZAMS fitting we derived 198 and 51 `physical' members with the probability $P\geq0.5$ for Berkeley 68 and Stock 20, respectively. Only these stars were analysed in the subsequent estimation of astrophysical parameters by the current paper. Using the {\sc UPMASK} method, \citet{Cantat-Gaudin20} obtained 200 and 38 member stars with the cluster membership probability greater than $P\geq0.7$ and $G$ magnitudes brighter than $G\leq18$ mag for Berkeley 68 and Stock 20, respectively. There is a general sense of agreement between the two studies. In the current study we limited the number of member stars considering both cluster $r_{\rm lim}$ radius  and also restricting the MS to stars with a cluster membership probability $P\geq0.5$. These differences are the likely cause for the different stellar counts between the two studies. Figs.~\ref{fig:cmds}a (for Berkeley 68) and \ref{fig:cmds}c (for Stock 20) are the $V\times (B-V)$ diagrams that show the distribution of field stars and the most probable members for each cluster along with the fitted ZAMS. Figs.~\ref{fig:cmds}b (for Berkeley 68) and \ref{fig:cmds}d (for Stock 20) show the location of the most probable cluster members on $G\times (G_{\rm BP}-G_{\rm RP})$ CMDs.

% FIGURE 4
\begin{figure*}[!t]
\centering
\includegraphics[scale=0.8, angle=0]{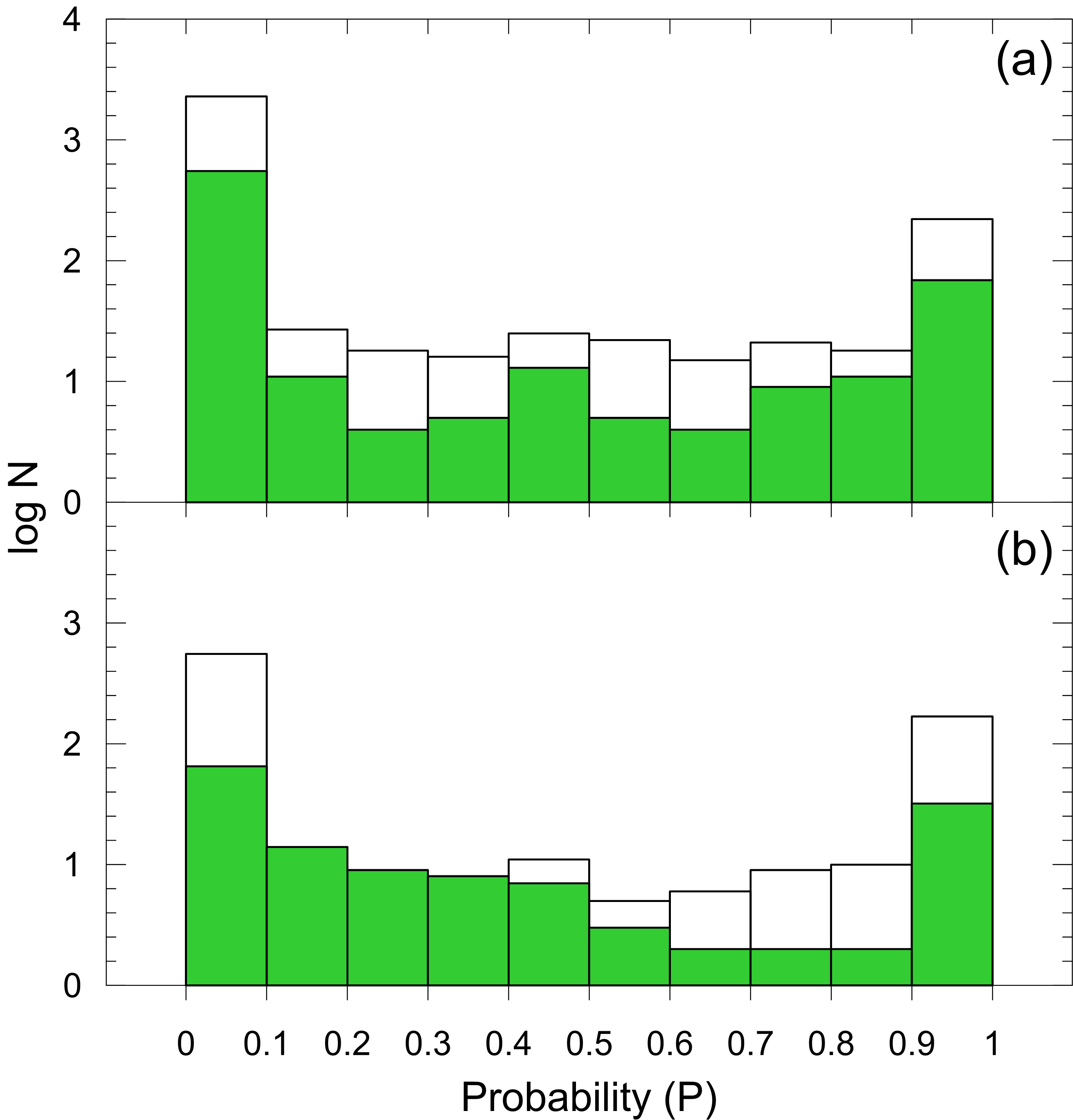}
\caption{Distribution for the membership probabilities of the stars in the  Berkeley 68 (a) and Stock 20 (b) regions. The white shading presents the stars that detected in the cluster areas, while the green colored shading denotes the stars that lie within the main-sequence band and cluster $r_{\rm lim}$ radii.
\label{fig:prob_hists} }
\end {figure*}

\begin{figure}[t]
\centering
\includegraphics[scale=.42, angle=0]{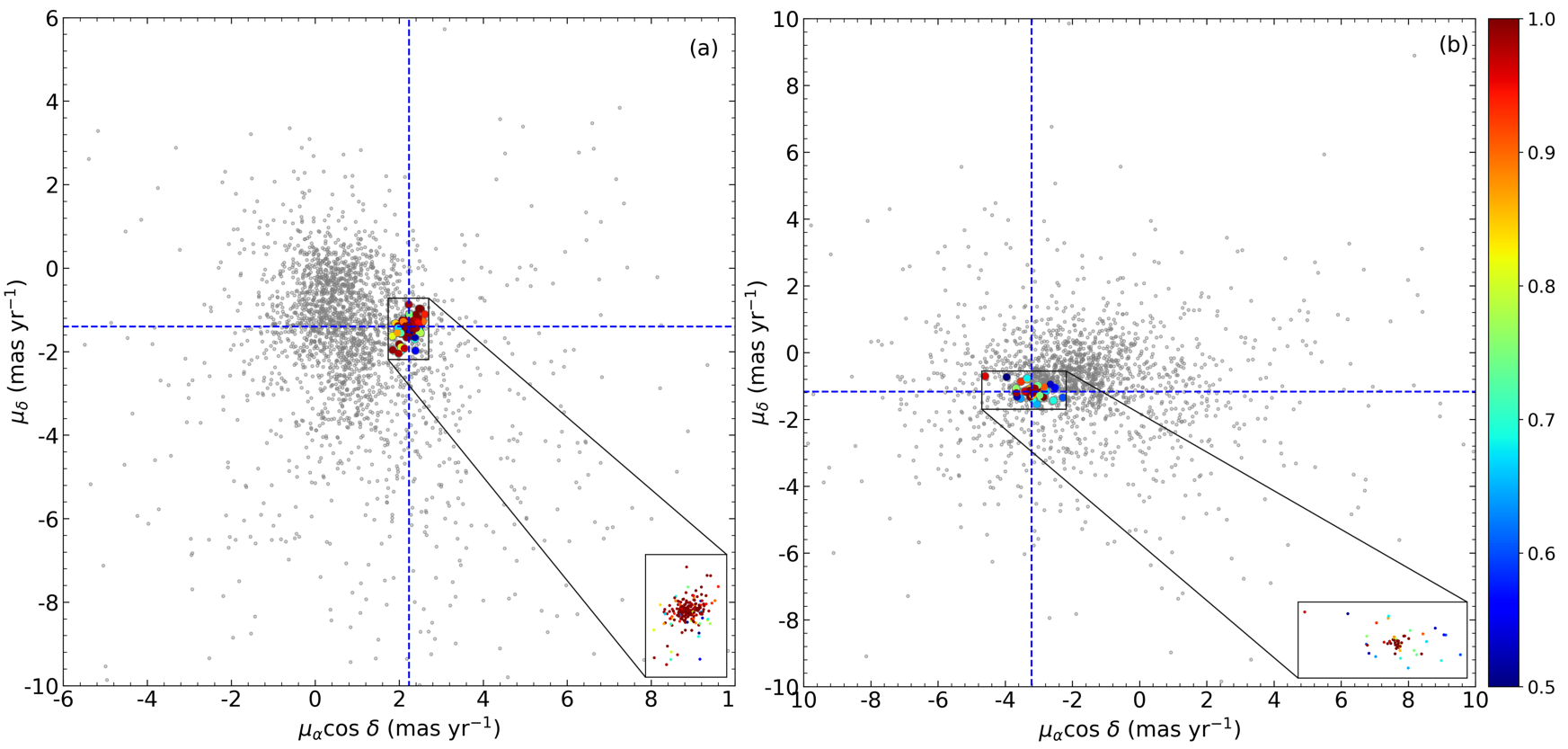}\\
\caption{VPDs of Berkeley 68 (a) and Stock 20 (b) based on {\it Gaia} DR3 astrometry. The membership probabilities of the stars are identified with the color scale shown on the right. The zoomed boxes in panels (a) and (b) show the region of condensation for each clusters in the VPDs. The intersection of the dashed blue lines are the point of mean proper motion values.
\label{fig:VPD_all}} 
 \end {figure}

\begin{figure}[t]
\centering
\includegraphics[scale=.42, angle=0]{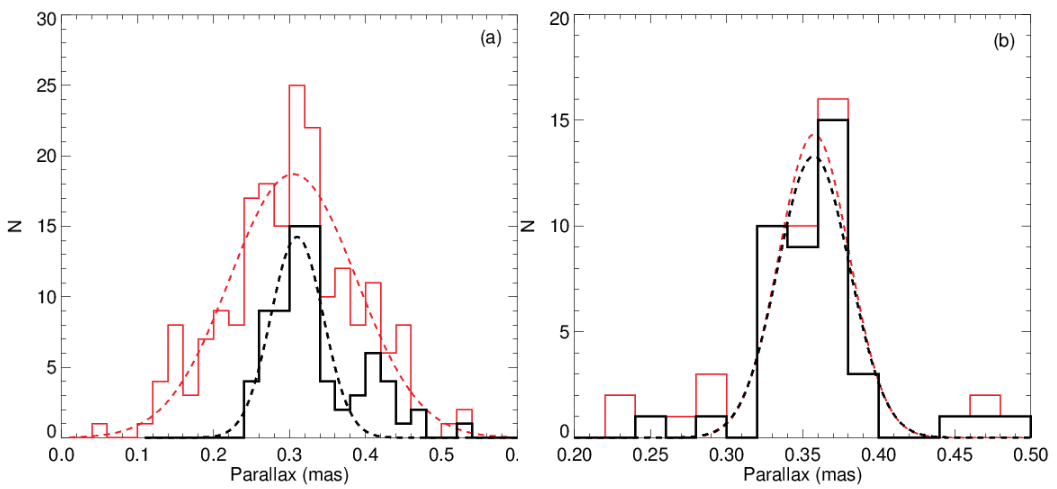}\\
\caption{{\it Gaia} DR3 based trigonometric parallax histograms for Berkeley 68 (a) and Stock 20 (b). The red-coloured histograms are based on stars selected without limiting  the relative parallax error and applied Gaussian fit, whereas the distribution of stars within the 0.2 relative parallax error and applied Gaussian fit are shown in black.
\label{fig:plx_hist}}
\end {figure}

The membership probabilities versus the number of stars detected through the two cluster regions are plotted and presented in Figure~\ref{fig:prob_hists}. To picture the positions of most probable member stars in each cluster we plotted vector-point diagrams (VPDs), shown as Fig.~\ref{fig:VPD_all}. It can be seen in the figure that Berkeley 68 (Fig.~\ref{fig:VPD_all}a) is more distinctly separated from field stars than Stock 20 is (Fig.~\ref{fig:VPD_all}b). In Fig.~\ref{fig:VPD_all}, the blue dashed lines indicate the mean proper motion values calculated from the most probable cluster members. These values were estimated as ($\mu_{\alpha}\cos\delta$, $\mu_{\delta}$)=($2.237\pm0.007$, $-1.401\pm0.005$) mas yr$^{-1}$ for Berkeley 68 and ($\mu_{\alpha}\cos\delta$, $\mu_{\delta}$)=($-3.215\pm0.004$, $-1.172\pm0.004$) mas yr$^{-1}$ for Stock 20. Mean proper motion component values calculated in the current study are compatible with the results of all studies performed with {\it Gaia} observations for the two clusters (see Table~\ref{tab:literature}). To calculate mean trigonometric parallaxes $\varpi$ of each cluster, we constructed the parallax histogram considering most probable cluster members and applied the Gaussian fit to the distribution (Fig.~\ref{fig:plx_hist}). During the calculation of $\varpi$, we considered the stars with a relative parallax error of less than 0.2 to minimize uncertainties. We obtained the mean $\varpi$ of Berkeley 68 as $0.31\pm 0.03$ mas and $0.36\pm 0.03$ mas for Stock 20. We also applied the linear equation of $d({\rm pc})=1000/\varpi$ (mas) to the mean trigonometric parallaxes and  obtained parallax distances as $d_{\varpi}=3226\pm 312$ pc and $d_{\varpi}=2778\pm 232$ pc for Berkeley 68 and Stock 20, respectively. To compare results, we calculated trigonometric parallaxes without limiting the relative parallax error values and obtained mean $\varpi$ as $0.30\pm 0.09$ mas and $0.36\pm 0.03$ mas for Berkeley 68 and Stock 20. It can be observed from the results that the uncertainty of $\varpi$ for Berkeley 68 is getting larger when we consider all relative parallax errors, whereas the results do not change for Stock 20. In Fig.~\ref{fig:plx_hist} red histograms and red dashed lines show the distribution of the most probable stars that were selected without limiting the relative parallax error and Gaussian fit, respectively, whereas the black solid lines present the histograms of stars with relative parallax error of less than 0.2. Gaussian fits are shown with black dashed lines. 

% FIGURE 5

% FIGURE 6

% FIGURE 07

%---------------------------------------------------------------
%\clearpage
\section{Astrophysical Parameters of the Clusters}
This section summarizes the procedures for the astrophysical analyses of Berkeley 68 and Stock 20. We determined the reddening and photometric metallicities separately using the two-color diagrams (TCDs). Keeping these two parameters as constant, we obtained the distance moduli and age of the clusters on CMDs simultaneously \citep[for detailed descriptions on the methodology see][]{Yontan15, Yontan19, Yontan21, Yontan22, Ak16, Bilir06, Bilir10, Bilir16, Bostanci15, Bostanci18}. 

%---------------------------------------------------------------
\subsection{Reddening}
\begin{figure}
\centering
\includegraphics[scale=0.5, angle=0]{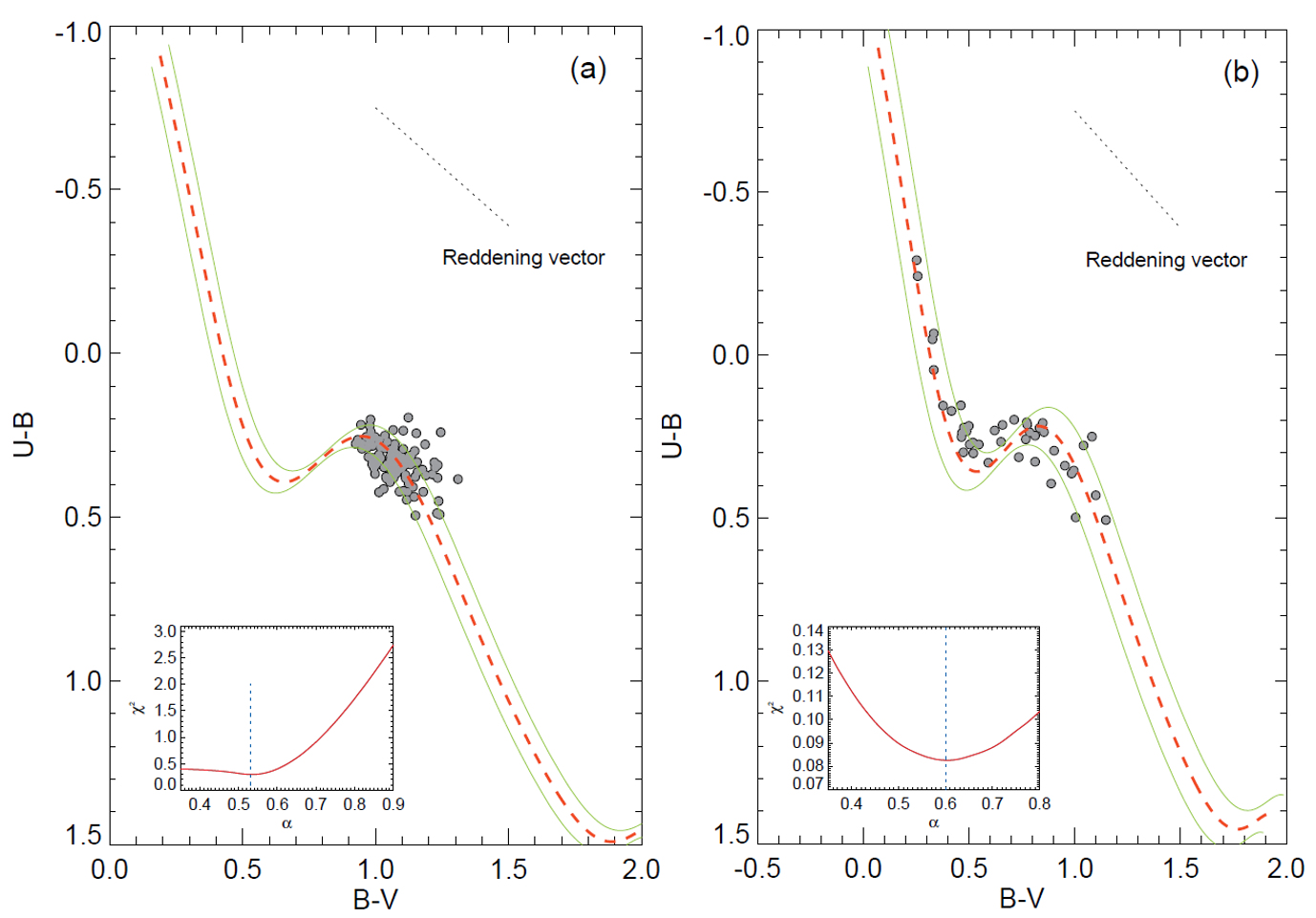}
\caption{TCDs of the most probable member main-sequence stars in the regions of clusters Berkeley 68 (a) and Stock 20 (b). Red dashed and green solid curves represent the reddened ZAMS given by \citet{Sung13} and $\pm1\sigma$ standard deviations, respectively. In the inner panels $\chi^2$ versus reddening vector values are shown. Dashed blue lines represent the best reddening vector corresponding $\chi^2_{\rm min}$. 
\label{fig:tcds}} 
\end{figure}

The separate estimation of the reddening in the direction of the cluster is important to obtain the precise age and distance of the studied cluster. In this study we used  $(U-B)\times (B-V)$ TCDs to determine color excesses across each cluster. To do this, we selected the most probable main-sequence stars ($P\geq 0.5$) within the magnitude ranges  $17\leq V \leq 19$ for Berkeley 68 and $12.75\leq V \leq 19$ mag for Stock 20. We constructed $(U-B)\times (B-V)$ diagrams of the most probable member stars and fitted the intrinsic solar-metallicity ZAMS of \citet{Sung13} to these. For $E(B-V)$ and $E(U-B)$, the ZAMS was shifted along different values of the reddening vector $\alpha=E(U-B)/E(B-V)$ utilizing $\chi^2$ minimization analyses. In the case of Berkeley 68, the best solution of minimum $\chi^2$ corresponds to the reddening vector being $\alpha=0.53$ and the color excess $E(B-V)=0.520\pm 0.032$ mag. For Stock 20 the best values were $\alpha=0.60$ for the reddening vector and $E(B-V)=0.400\pm 0.048$ mag for the color excess. In Fig.~\ref{fig:tcds}, we present the TCDs with the best results. The derived errors for the color excesses are the $\pm 1\sigma$ deviations. The estimated color excess for Berkeley 68 is in good agreement with the value $E(B-V)=0.52\pm 0.04$ mag given by \citet{Maurya20}. It is slightly lower than $E(B-V)=0.67$ mag estimated by \citet{Kharchenko12} and greater than the $E(B-V)=0.445$ mag of \citet{Cantat-Gaudin20}. Moreover, the derived color excesses for Stock 20 agrees well with the $E(B-V)=0.352$ and $E(B-V)=0.48\pm 0.01$ mag given by \citet{Cantat-Gaudin20} and \citet{Dias21}, respectively, whereas it is higher than the value of $E(B-V)=0.2$ mag represented by \citet{Kharchenko12} (see Table~\ref{tab:literature} for detailed comparison).
%---------------------------------------------------------------

% FIGURE 08

% FIGURE 09
\begin{figure*}
\centering
\includegraphics[scale=0.6, angle=0]{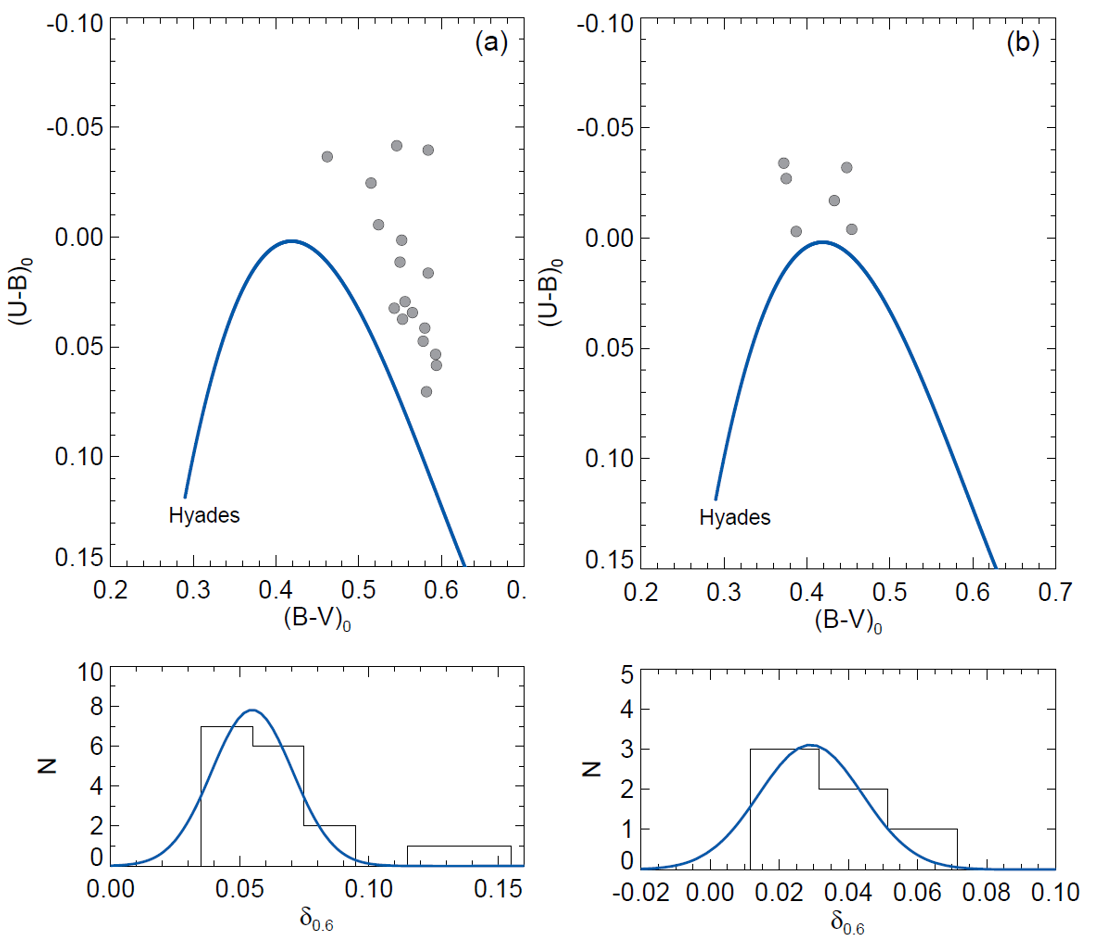}
\caption{$(U-B)_0\times(B-V)_0$ TCDs (upper panels) and histograms for the normalized $\delta_{0.6}$ (lower panels) for 17 (Berkeley 68, panel a) and 6 (Stock 20, panel b) most probable F-G type main sequence stars. The solid blue lines in the TCDs represent the main-sequence of Hyades and Gaussian fits, respectively.
\label{fig:hyades}} 
\end {figure*}

%---------------------------------------------------------------
\newpage
\subsection{Photometric Metallicity}
\label{sec:metallicity}

Applying \citet{Karaali11}'s method to the TCDs of Berkeley 68 and Stock 20, the photometric metallicity [Fe/H] was determined using the mean UV-excess ($\delta_{0.6}$) of the most probable ($P\geq 0.5$) F- and G-type main-sequence stars within the color index of $0.3\leq (B-V)_0\leq0.6$ mag \citep{Eker18, Eker20}. We calculated de-reddened $(B-V)_0$ and $(U-B)_0$ colors of most probable main-sequence stars and limited their $(B-V)_0$ color to the range $0.3\leq (B-V)_0\leq0.6$  to select F- and G- type main-sequence stars for the analyses. This resulted in the selection of 17 stars for Berkeley 68 and 6 stars for Stock 20. We plotted the $(U-B)_0\times(B-V)_0$ TCDs for the selected stars and the Hyades main-sequence to compare $(U-B)_0$ colors of the data sets with the same de-reddened $(B-V)_0$ color and calculate the differences between the $(U-B)_0$ color indices. These differences are defined as UV-excess ($\delta$) which is given by the equation $\delta =(U-B)_{\rm 0,H}-(U-B)_{\rm 0,S}$, where H and S subscripts represent the Hyades and cluster stars respectively.

To employ the method detailed in \citet{Karaali11}, we normalised the derived UV-excess of the selected F-G type stars to the UV-excess at $(B-V)_0 = 0.6$ mag (i.e. $\delta_{0.6}$) and calculated the mean $\delta_{0.6}$ via fitting Gaussians to the constructed $\delta_{0.6}$ histograms \citep{Karaali03a, Karaali03b}. The peak of the Gaussian gives the mean $\delta_{0.6}$ value which was determined as $\delta_{0.6}=0.055\pm0.016$ mag for Berkeley 68 and $\delta_{0.6}=0.029\pm0.015$ mag for Stock 20. Here, the uncertainty of the mean $\delta_{0.6}$ is the $\pm1\sigma$ standard deviation of the Gaussian fit. The photometric metallicity of studied clusters were determined by evaluating mean $\delta_{0.6}$ value in the equation given by \citet{Karaali11}:   
\begin{eqnarray}
{\rm [Fe/H]}=-14.316(1.919)\delta_{0.6}^2-3.557(0.285)\delta_{0.6}+0.105(0.039).
\end{eqnarray}
$(U-B)_0\times(B-V)_0$ TCDs and histograms of $\delta_{0.6}$ for the selected stars in two clusters are given as Fig.~\ref{fig:hyades}. The photometric metallicities estimated from the most likely cluster members are [Fe/H]=$-0.13\pm 0.08$ for Berkeley 68 and [Fe/H]=$-0.01\pm 0.06$ dex for Stock 20. 

We transformed the photometric  metallicity [Fe/H] to the mass fraction $z$ for the selection of isochrones and hence age determination. For transformation between the two parameters we used the analytic equations provided by Bovy\footnote{https://github.com/jobovy/isodist/blob/master/isodist/Isochrone.py} which are developed for {\sc parsec} isochrones \citep{Bressan12}. The relevant equations are: 

\begin{equation}
z_{\rm x}={10^{{\rm [Fe/H]}+\log \left(\frac{z_{\odot}}{1-0.248-2.78\times z_{\odot}}\right)}}
\end{equation}      
and
\begin{equation}
z=\frac{(z_{\rm x}-0.2485\times z_{\rm x})}{(2.78\times z_{\rm x}+1)}.
\end{equation} 
Here $z_{\rm x}$ and $z_{\odot}$ are intermediate values where solar metallicity $z_{\odot}$ was adopted as 0.0152 \citep{Bressan12}. We calculated $z=0.012$ for Berkeley 68 and $z=0.015$ for Stock 20.

%-----------------------------------------------------------

\subsection{Distance Moduli and Age Estimation}

CMDs were constructed using values for the most probable member stars.  The morphology of these CMDs allows identification of the main-sequence, turn-off point, and giant members.  It leads to a model-based mass, age, and distance for each star \citep{Bisht19}. In this study we estimated the distance moduli and age simultaneously by fitting {\sc parsec} isochrones of \citet{Bressan12} to the $UBV$ and {\it Gaia} based CMDs of the two clusters. 

\begin{figure*}
\centering
\includegraphics[scale=.60, angle=0]{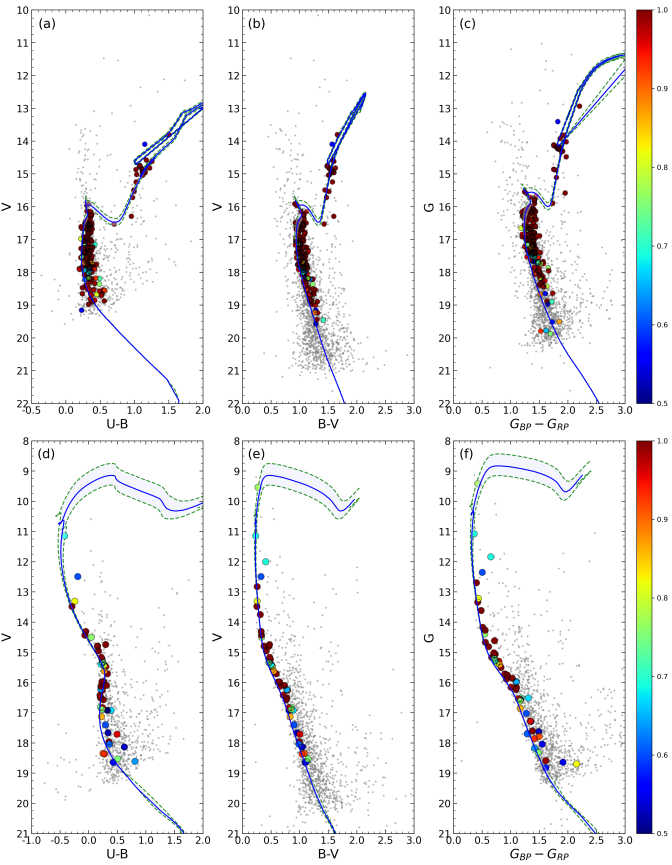}
\caption{$U\!BV$ and {\it Gaia} based CMDs for the Berkeley 68 (panels a, b, and c) and Stock 20 (panels d, e, and f). Membership probabilities of most probable cluster stars are represented with different color scales that are shown on the colorbars to the right, whereas field stars are shown with grey colored dots. The best fitting {\sc parsec} isochrones and their errors are presented as the blue lines with the surrounding purple shaded areas. Superimposed isochrone ages match to 2.4 Gyr for Berkeley 68 and 50 Myr for Stock 20.
\label{fig:figure_ten} }
\end {figure*}

To determine the distance moduli and age of the clusters, we constructed $V\times (U-B)$, $V\times (B-V)$, and $G\times (G_{\rm BP}-G_{\rm RP})$ diagrams and visually fitted the isochrones \citep{Bressan12} considering the most probable main-sequence, turn-off and giant members ($P\geq 0.5$). {\sc Parsec} isochrones were selected according to the mass fraction $z$ obtained above (Section~\ref{sec:metallicity}). The $U\!BV$ data isochrone fitting procedure made reference to $E(B-V)$ values calculated by this study. For {\it Gaia} DR3 data we used the expression $E(G_{\rm BP}-G_{\rm RP})= 1.41\times E(B-V)$ presented in the study of \citet{Sun21}. Error estimations for distance moduli and distances considered a relation given by \citet{Carraro17}. Uncertainties in age were determined using low and high age isochrones whose values fitted the observed scatter about the MS. Fig.~\ref{fig:figure_ten} presents the $V\times (U-B)$, $V\times (B-V)$ and $G\times (G_{\rm BP}-G_{\rm RP})$ CMDs with best fitting isochrones. The derived distance moduli and ages for the two clusters are given as follow:

\begin{itemize}
\item{{\bf Berkeley 68}: Isochrones of different ages ($\log({\rm age}) = 9.34, 9.38$ and 9.41 with $z =0.012$) were superimposed onto the $V\times (U-B)$, $V\times (B-V)$, and $G\times (G_{\rm BP}-G_{\rm RP})$ diagrams. Considering only the most probable main-sequence stars, we derived the distance modulus as $\mu_{V}=14.00\pm 0.12$ mag. This leads to the isochrone distance to be $d_{\rm iso}=3003\pm 165$ pc.  If we apply over-weights to the turn-off point and giant member stars of the cluster, the age is estimated as $t=2.4\pm 0.2$ Gyr. The derived isochrone distance is compatible with the values given by \citet{Cantat-Gaudin18}, \citet{Tarricq21} and \citet{Dias21} within the errors (see Table~\ref{tab:literature}). The isochrone distance of Berkeley 68 is compatible with the {\it Gaia} DR3 trigonometric parallax distance value $d_{\varpi}=3226\pm 312$ pc estimated in this study. Also, considering the estimated age and its error in the study, it can be concluded that the value is compatible with the results represented in the studies of \citet{Cantat-Gaudin18} and \citet{Tarricq21} (see Table~\ref{tab:literature} for detailed comparison)}.

\item{{\bf Stock 20}: The same methodology was applied to this cluster, using isochrones for $\log({\rm age}) = 7.60, 7.70$ and 7.78 with $z = 0.015$. The distance modulus is $\mu_{\rm V}=13.56\pm 0.16$ mag, corresponding to an isochrone-based distance of  $d_{\rm iso}=2911\pm 216$ pc. Cluster age is  $t=50\pm 10$ Myr. Because of its young age, we took into account the main-sequence members during distance moduli and age estimation. The derived distance is compatible with the value $d=2600$ pc given by \citet{Buckner13}, as well as with the results found by most researches whose results are listed in Table~\ref{tab:literature}. The estimated age of the cluster is lower than the values of $t=220$ Myr of \citet{Kharchenko12} and \citet{Hao21}, whereas higher than the value $t=8\pm 1$ Myr of \citet{Liu19} (see Table~\ref{tab:literature} for detailed comparison). The isochrone distance is in good agreement with the value $d_{\varpi}=2778\pm 232$ pc obtained by the current study from {\it Gaia} DR3 trigonometric parallaxes.}

\end{itemize}
\noindent
The distances determined from the isochrone fitting for the clusters are similar to those derived from the {\it Gaia} DR3 trigonometric parallaxes.

% FIGURE 10

%---------------------------------------------------------------

\section{Kinematics and Galactic Orbits Parameters}
To obtain the galactic orbital parameters for the two clusters, we used the Python language based galactic dynamics library \citep[{\sc galpy}\footnote{See also https://galpy.readthedocs.io/en/v1.5.0/},][]{Bovy15} in which the potential functions are defined. Orbital integrations utilized via {\sc MWPotential2014} \citep{Bovy15} which assumes an axisymmetric potential for the Milky Way galaxy. The {\sc MWPotential2014} models a three-component model of the Galaxy consisting of bulge, disc, and halo potentials. The bulge component is described as a spherical power-law density profile represented in \citet{Bovy15}. The galactic disc potential is defined by Miyamoto–Nagai expressions \citep{Miyamoto75}. Considering the spherically symmetric spatial distribution of dark matter in the halo, the halo potential is defined by the Navarro Frank–White profile \citep{Navarro96}. For orbital analyses we took into account the parameters which are listed for the three components in \citet[][see Table 1 of that paper]{Bovy15}. The Sun's galactocentric distance and orbital velocity were taken as $R_{\rm GC}=8$ kpc and $V_{\rm rot}=220$ km s$^{-1}$, respectively \citep{Bovy15, Bovy12}. The Sun's distance from the galactic plane was assumed to be $27 \pm 4$ pc \citep{Chen00}.   

For successful integration of the orbit,  one of the necessary parameters to know is radial velocity of the clusters. We adopted the radial velocity value for Berkeley 68 cluster as calculated by \citet{Soubiran18}, but did not have a corresponding value for Stock 20. Therefore, we estimated the velocity components and galactic orbital parameters for only Berkeley 68. To do this, we used the {\sc MWPotential2014} code with the following inputs: the central equatorial coordinates ($\alpha=04^{\rm h} 44^{\rm m} 13^{\rm s}$, $\delta= +42^{\rm o} 08^{\rm '} 02^{\rm''}$) adopted from \citet{Cantat-Gaudin20}, the mean proper motion components ($\mu_{\alpha}\cos\delta$, $\mu_{\delta}$=$2.237\pm0.007$, $-1.401\pm0.005$ mas yr$^{-1}$) obtained in Section 3.3, isochrone distance ($d_{\rm iso}=3003\pm 165$ pc) from Section 4.3, and the radial velocity ($V_{\rm r}=-20.31\pm 1.86$ km s$^{-1}$) given by \citet{Soubiran18}. See also Table~\ref{tab:Final_table} on page \pageref{tab:Final_table}. 

We integrated the orbit forwards and backwards in time with 1 Myr steps up to the age 2.5 Gyr in order to estimate its possible current location and birth place. We did not integrate the orbit more than the cluster age given that inaccuracies of the used potentials depend on time and so influence the reliability of the kinematic and dynamic analyses results. Additional errors of the input parameters (proper motion components, distance and radial velocity) will affect the orbit determination, so the results are approximate \citep{Tarricq21, Sariya21}. 

Orbital integration resulted in the following estimates for Berkeley 68: apogalactic ($R_{\rm a}=10901\pm149$ pc) and perigalactic ($R_{\rm p}=8893\pm34$ pc) distances, eccentricity ($e=0.101\pm0.005$), maximum vertical distance from galactic plane ($Z_{\rm max}=470\pm33$ pc), space velocity components ($U,V,W$=$+7.79\pm2.41$, $-40.27\pm1.31$, $11.92\pm0.65$ km s$^{-1}$), and orbital period ($P_{\rm orb}=284\pm3$ Myr). \citet{Soubiran18} used {\it Gaia} DR2 \citep{Gaia18} astrometric data, finding the space velocity components of Berkeley 68 to be $(U, V, W)$=($6.97\pm1.78$, $-42.45\pm0.92$, $14.46\pm0.32$ km s$^{-1}$) which is in good agreement with the values calculated in this study. We applied a local standard of rest (LSR) correction to the $(U, V, W)$ space velocity components considering the values ($8.83\pm0.24$, $14.19\pm0.34$, $6.57\pm0.21$) km s$^{-1}$ of \citet{Coskunoglu11}. The LSR corrected space velocity components are $(U, V, W)_{\rm LSR}$ = ($16.62\pm2.42$, $-26.08\pm1.35$, $18.49\pm0.69$) km s$^{-1}$. Using these LSR corrected space velocities, we calculated the total space velocity as $S_{\rm LSR}=36.03\pm2.85$ km s$^{-1}$. This result is agreement for the velocity of young thin-disc stars \citep{Leggett92}.

\begin{figure*}
\centering
\includegraphics[scale=0.8, angle=0]{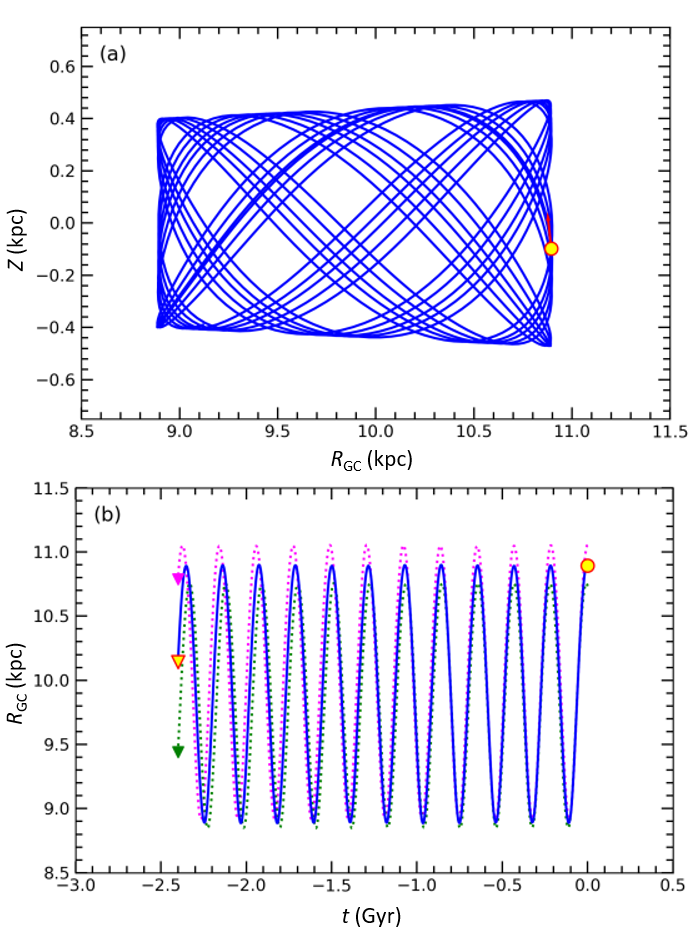}
\caption{\label{fig:galactic_orbits}
The orbit of Berkeley 68 in the $Z \times R_{\rm GC}$ and $R_{\rm GC} \times t$  planes. The filled yellow circles and triangle are the present day and birth positions, respectively. The red arrow shows the motion vector of the cluster. The pink and green dotted lines show the orbit when errors in input parameters are considered, while the pink and green filled triangles represent the birth locations of the cluster based on the upper and lower error estimates.} 
\end {figure*}

According to the results, Berkeley 68 intersects the galactic plane around eight times during its orbital period,  rising up to maximum $Z_{\rm max}=470\pm33$ pc above the galactic plane.  This indicates that Berkeley 68 belongs to the thin-disc component of the Galaxy. The orbit of the cluster is slightly eccentric with the value of $e=0.101\pm0.005$. The birth radius of the Berkeley 68 is calculated as 10.16$^{+0.65}_{-0.71}$ kpc (see Fig.~\ref{fig:galactic_orbits}b). This radius is outside the solar circle which shows that Berkeley 68 was born in a metal-poor region. Investigating $R_{\rm p}$ and $R_{\rm a}$ distances, we can conclude that Berkeley 68 completely orbits outside the solar circle (see Fig.~\ref{fig:galactic_orbits}a). 

Berkeley 68's orbits on the $Z \times R_{\rm GC}$ and $R_{\rm GC} \times t $ planes are represented in Fig.~\ref{fig:galactic_orbits}. The motion of Berkeley 68 is shown as a function of distance from the galactic center and the galactic plane as the `side view' of Figure~\ref{fig:galactic_orbits}a. The red arrow shows the direction of motion for Berkeley 68. Yellow filled circles are the present-day location of the cluster (Figs.~\ref{fig:galactic_orbits}a and \ref{fig:galactic_orbits}b), whereas the yellow filled triangle represents the birth position of the cluster (Fig.~\ref{fig:galactic_orbits}b). Figure~\ref{fig:galactic_orbits}b represents the distance of the cluster on the galactic plane with time. The pink and green dotted lines show the cluster movement in time when the upper and lower errors in proper motion, radial velocity and distance are considered. The pink and green filled triangles represent the position of the cluster's birthplace. As can be seen from Fig.~\ref{fig:galactic_orbits}b, different input parameters can cause the estimated birth locations of clusters to change. In the $R_{\rm GC} \times t $ plane of figure Berkeley 68 has an uncertainty of about 1.5 kpc for its possible birthplace. 

We interpreted that the relative errors in $R_{\rm p}$ and $R_{\rm a}$ do not exceed 2\%, whereas the relative error for $Z_{\rm max}$ parameter is about 7\%, which lends us confidence that the orbital solutions are reasonable.

% FIGURE 11

%---------------------------------------------------------------

\section{Dynamical Study of the Clusters}

\subsection{Luminosity Functions}

\begin{figure}
\centering
\includegraphics[scale=1.2, angle=0]{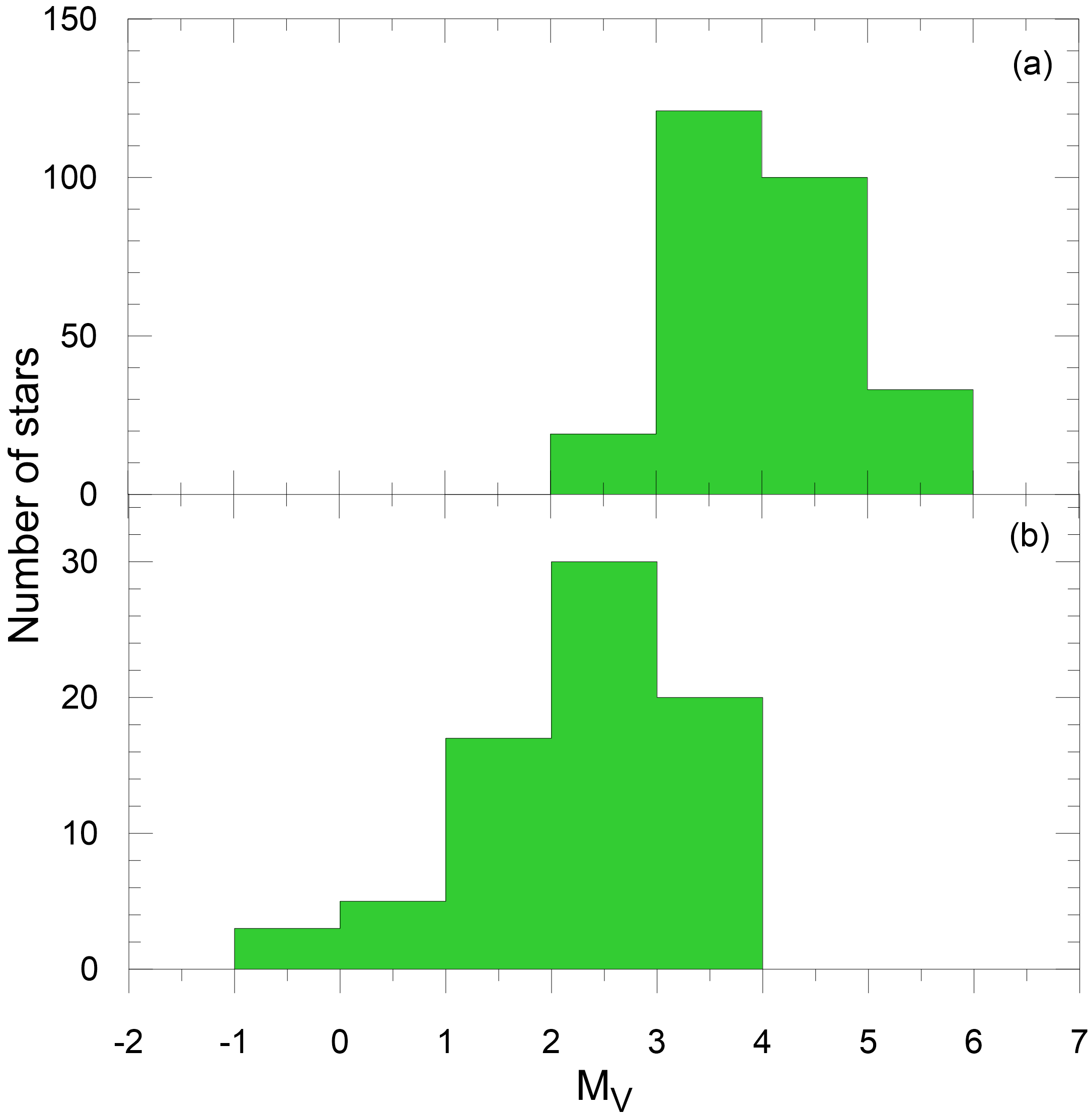}
\caption{\label{fig:luminosity_functions}
The LFs based on selected stars for Berkeley 68 (a) and Stock 20 (b).}
\end {figure}

The luminosity functions (LF) and mass functions (MF) of OCs are related via well known mass-luminosity relationships. The LF for an open cluster is described as the distribution of the relative number of main-sequence members in different absolute magnitude ranges. To measure the LFs of the studied clusters, we considered main-sequence stars with membership probability $P>0$ located inside the previously obtained limiting radii of the clusters. For Berkeley 68, apparent magnitudes of selected main-sequence stars inside the $r_{\rm lim}=8'$  radius lie within the range $16.75 \leq V \leq 20$ mag, and for Stock 20 apparent magnitudes of selected stars located inside the $r_{\rm lim}=7'.5$ radius are within $13 \leq V \leq 17$ mag. We converted apparent $V$ magnitudes of the selected stars to their absolute $M_{\rm V}$ magnitudes using the distance modulus expression $M_{\rm V} = V-5\times \log d +3.1\times E(B-V)$. $V$, $d$ and $E(B-V)$ are apparent magnitude, isochrone distance, and color excess as previously estimated for each cluster, see also Table~\ref{tab:Final_table}. Via these calculations, we achieved the $2< M_{\rm V}< 6$ and $-1< M_{\rm V}< 4$ mag absolute magnitude ranges of the selected stars for Berkeley 68 and Stock 20 respectively. The resulting LF histograms were constructed with a bin width of 1 mag and are presented in Fig.~\ref{fig:luminosity_functions} for each cluster. It can be seen from the figure that the LFs rise up to around $M_{\rm V} = 3.5$ mag for Berkeley 68 (Fig.~\ref{fig:luminosity_functions}a) and $M_{\rm V} =2.5$ mag for Stock 20 (Fig.~\ref{fig:luminosity_functions}b), respectively.     

% FIGURE 12

%---------------------------------------------------------------
\subsection{Mass Functions}

The initial mass function (IMF) is an empirical function that defines the initial distribution of masses per unit volume for a population of stars at the time of their formation. Investigating the IMF is useful to understand the ensuing evolution of the cluster. However, a direct determination of the IMF is difficult due to the evolution of cluster stars which takes place at different rates for different stellar masses \citep{Joshi22}. From the present distribution of stellar masses we can measure the present day mass function (PDMF) for the clusters. The PDMF is associated with the LF according to a mass-luminosity relation, and is expressed by the following equation:

\begin{eqnarray}
{\rm log(dN/dM)}=-(1+\Gamma)\times \log M + {\rm constant}.
\label{eq:mass_luminosity}
\end{eqnarray}
Here $dN$ is the number of stars per unit mass $dM$, the central mass is designated as $M$, and $\Gamma$ is the slope of the PDMF.  

\begin{figure}
\centering
\includegraphics[scale=1.2, angle=0]{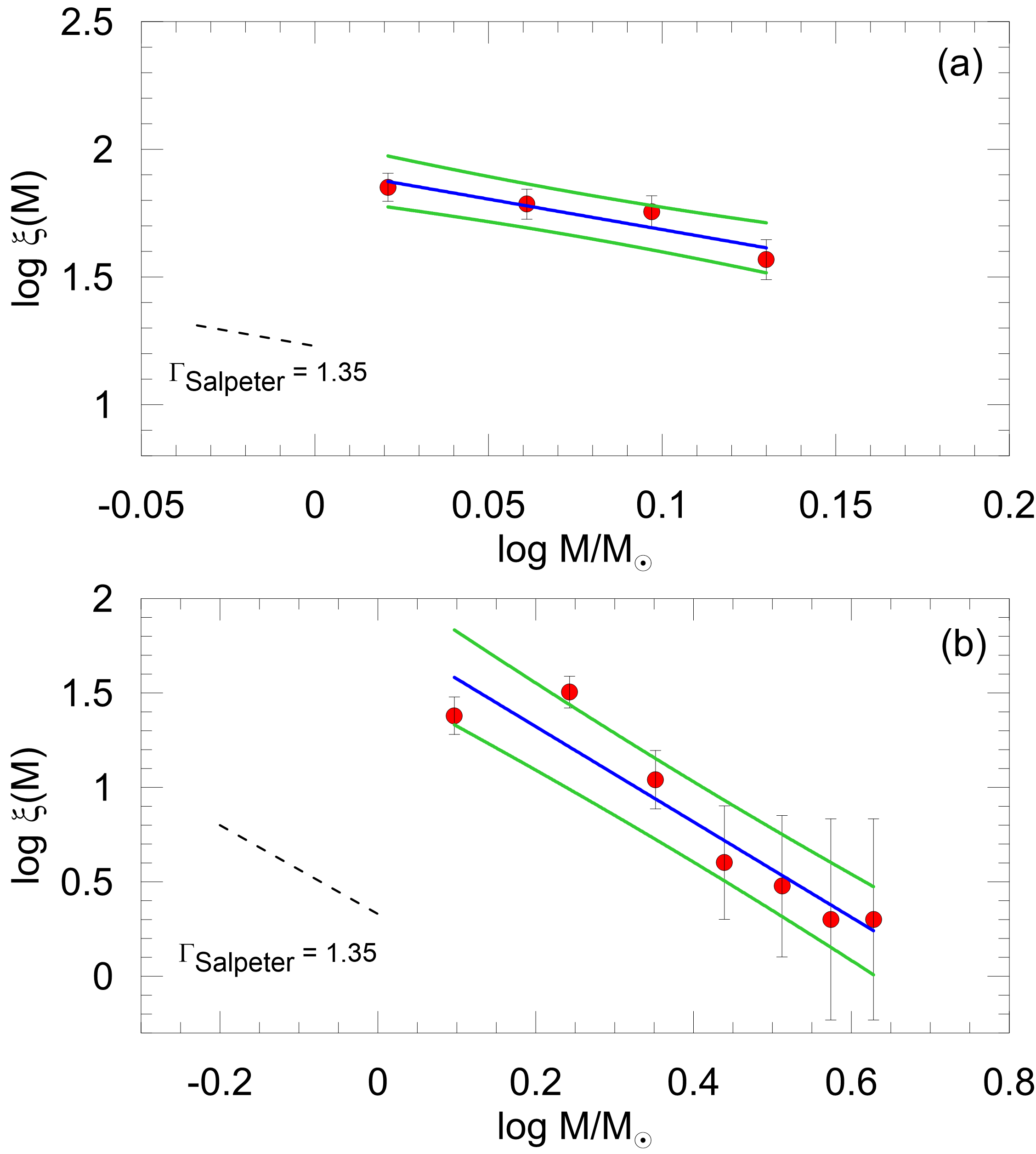}
\caption{\label{fig:mass_functions}
Present day mass functions of Berkeley 68 (a) and Stock 20 (b). Blue lines indicate the computed mass functions of the OCs, while green lines are the $\pm1\sigma$ standard deviations.}
\end {figure}

To obtain PDMFs we considered the {\sc parsec} isochrones according to the ages and metallicity fractions ($z$) obtained above for the clusters. Using these isochrones we transformed the LF into the PDMF, through application of the theoretical models of \citet{Bressan12} which produced a high degree polynomial equation between the $V$-band absolute magnitudes and masses. Using this absolute magnitude-mass relation we converted the observational absolute magnitudes $M_{\rm V}$ to masses. The number of stars with membership probability $P>0$ used in transformation processes are 273 for Berkeley 68 and 75 for Stock 20. The mass ranges of these stars were determined within $1\leq M/ M_{\odot}\leq 1.4$ for Berkeley 68 and $1.3\leq M/ M_{\odot}\leq 4.1$ for Stock 20. Using equation~\ref{eq:mass_luminosity}, we computed the PDMF slope values as $\Gamma=1.38 \pm 0.71$ for Berkeley 68 and as $\Gamma=1.53 \pm 0.39$ for Stock 20. PDMFs for both clusters are shown in Fig.~\ref{fig:mass_functions} 

% FIGURE 13

% FIGURE 14
\begin{figure}
\centering
\includegraphics[scale=0.80, angle=0]{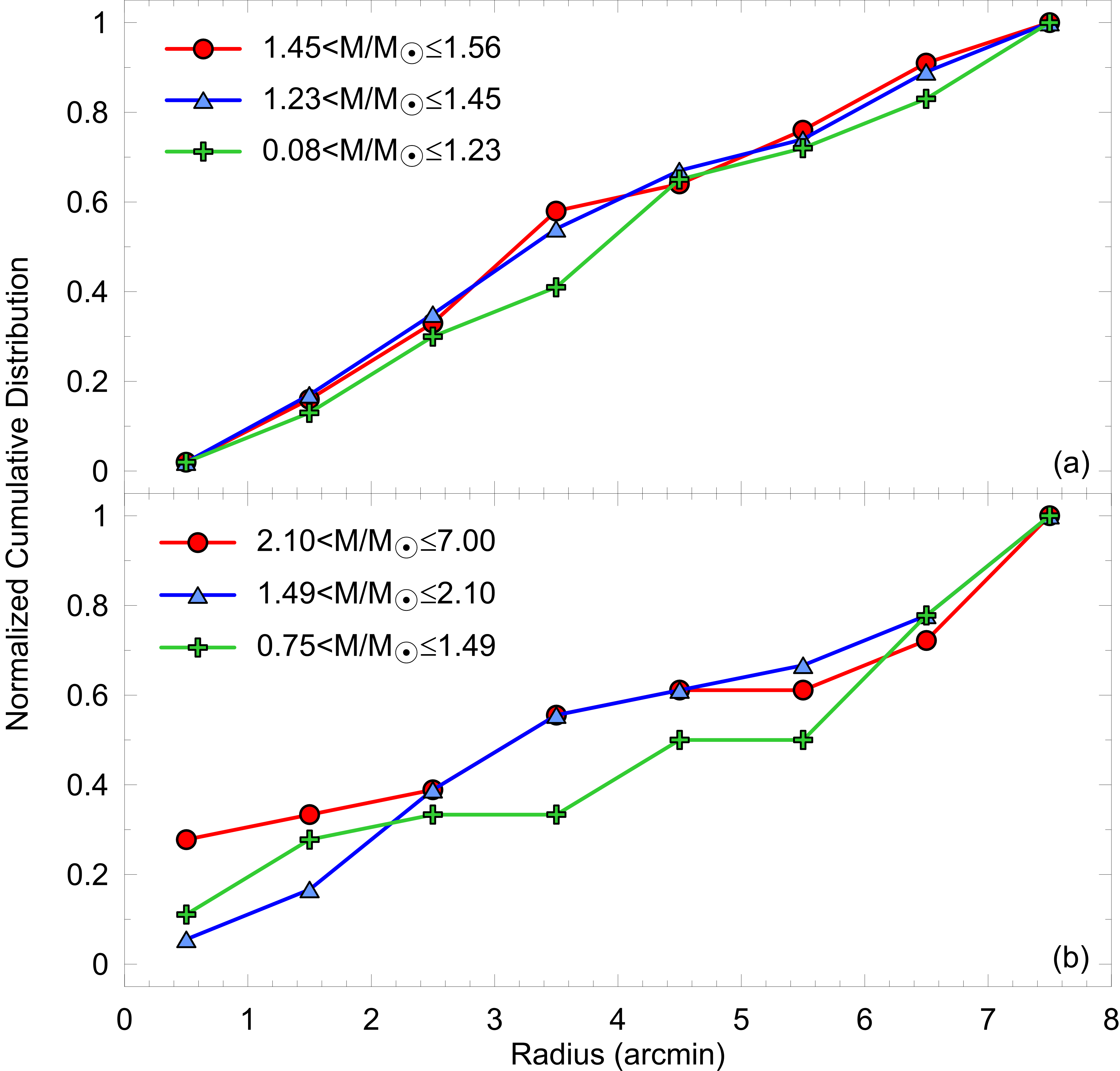}
\caption{\label{fig:radial_distributions}
The cumulative radial distribution of stars in different mass ranges for Berkeley 68 (a) and Stock 20 (b).}
\end {figure}

\subsection{The Dynamical State of Mass Segregation}

Mass segregation could occur at cluster formation. Hence investigation of mass segregation in young clusters could give insights into the process of cluster formation \citep{Sagar88, Raboud98, Fischer98, Alcock19}. Interactions between cluster stars can cause energy transfer from high-mass stars to low-mass stars. This  leads to more massive stars gradually moving toward the cluster center while lower-mass stars move toward the cluster halo, i.e., mass segregation \citep[see e.g.,][]{Hillenbrand98, Baumgardt03, Dib19, Bisht20, Pavlik22}. This energy exchange enables kinetic energy equipartition leading to the velocity distribution of cluster stars conforming to a Maxwellian distribution. The time scale at which this duration occurs is defined as relaxation time ($T_{\rm E}$). It is related to  dynamical evolution of the cluster. Relaxation time is defined by the following expression given by \citet{Spitzer71}:

\begin{equation}
T_{\rm E} = \frac{8.9 \times 10^{5} N^{1/2} R_{\rm h}^{3/2}}{\langle m\rangle^{1/2}\log(0.4N)}
\end{equation} 
where $N$ is the number of considered stars, $R_{\rm h}$ is the half-mass radius (in parsecs) of the cluster, and $\langle m \rangle$ is the average mass of the stars. 

The stars used in this study's calculations of relaxation times were selected by limiting their mean proper motion and trigonometric parallax values as per Section~3.3. Stars were selected by defining a 1 mas yr$^{-1}$ circular radius centered around the mean cluster proper motion values together with considering the upper and lower errors of trigonometric parallax for each cluster. With these limitations, 137 stars in Berkeley 68 and 54 stars in Stock 20 were within the probabilities $P>0$ in the mass range $0.08 < M/M_{\odot} \leq 1.56$ and $0.75 <M/M_{\odot}\leq 7$. The total mass of Berkeley 68 was estimated to be $M = 180.73\, M_{\odot}$ and for Stock 20 $M = 118.42 \, M_{\odot}$. This means that the average stellar mass is $\langle m \rangle = 1.32 M/M_{\odot}$ for Berkeley 68 and $\langle m \rangle = 2.19 M/M_{\odot}$ for Stock 20. Half-mass radii are  $R_{\rm h}$= 3.39 pc for Berkeley 68 and $R_{\rm h}$= 3.66 pc for Stock 20. The dynamical relaxation time of Berkeley 68 was estimated as $T_{\rm E}=32.55$ Myr and for Stock 20 as $T_{\rm E}=23.17$ Myr. We concluded that both clusters are dynamically relaxed due to the derived $T_{\rm E}$ ages being younger than the present ages of two clusters as estimated in the current study (see Table~\ref{tab:Final_table}).

To understand the impact of mass segregation effect in the two clusters, we divided the masses of selected stars into three ranges containing equal number of stars. These ranges are $0.08 < M/M_{\odot}\leq 1.23$ (high-mass), $1.23 < M/M_{\odot}\leq 1.45$ (intermediate-mass), and $1.45 < M/M_{\odot}\leq 1.56$ (low-mass) for Berkeley 68 and $2.10 < M/M_{\odot}\leq 7$, $1.49 < M/M_{\odot}\leq 2.10$, and $0.75 < M/M_{\odot}\leq 1.49$ for Stock 20. The normalized cumulative radial distributions of stars in these different mass ranges are shown in Fig.~\ref{fig:radial_distributions}. Fig.~\ref{fig:radial_distributions}a exhibits that the radial stellar distributions are similar for all three mass groups for Berkeley 68.  Hence, we suggest that the mass segregation is not occurring in Berkeley 68 (Fig.~\ref{fig:radial_distributions}a), while Stock 20 shows a mass segregation effect due to the massive stars being more centrally concentrated than the intermediate and low-mass (Fig.~\ref{fig:radial_distributions}b). We utilized the Kolmogorov–Smirnov (K-S) test, finding the confidence level for a mass segregation effect to be 62\% for Berkeley 68  and 97\% for Stock 20. 

% Table 7
\begin{table}
  \centering
  \renewcommand{\arraystretch}{0.75}
  \caption{Fundamental parameters of Berkeley 68 and Stock 20.}
  \medskip
  {\small
        \begin{tabular}{lrr}
\hline
Parameter & Berkeley 68 & Stock 20 \\
\hline
($\alpha,~\delta)_{\rm J2000}$ (Sexagesimal)& 04:44:12.72, $+$42:08:02.40 & 00:25:16.33, $+62$:37:26.40\\
($l, b)_{\rm J2000}$ (Decimal)              & 162.0391, $-$2.4041         & 119.9291, $-00$.0951       \\
$f_{0}$ (stars arcmin$^{-2}$)               & 8.204 $\pm$ 2.008           & 27.197 $\pm$ 7.164         \\
$f_{\rm bg}$ (stars arcmin$^{-2}$)          & 9.139 $\pm$ 0.402           & 13.742 $\pm$ 0.222         \\
$r_{\rm c}$ (arcmin)                        & 2.663 $\pm$ 0.616           & $0.543 \, \pm$ 0.225       \\
$r_{\rm lim}$ (arcmin)                      & 8                           & 7.5                        \\
$r$ (pc)                                    & 6.99 $\pm$ 0.38             & 6.35 $\pm$ 0.47            \\
Cluster members ($P\geq0.5$)                & 198                         & 51                         \\
$\mu_{\alpha}\cos \delta$ (mas yr$^{-1}$)   & +2.237 $\pm$ 0.007          & $-3.215 \, \pm$ 0.004      \\
$\mu_{\delta}$ (mas yr$^{-1}$)              & $-1.401 \, \pm$ 0.005       & $-1.172 \, \pm$ 0.004      \\
$\varpi$ (mas)                              & 0.31 $\pm$ 0.03             & 0.36 $\pm$ 0.03            \\
$d_{\varpi}$ (pc)                           & 3226 $\pm$ 312              & 2778 $\pm$ 232             \\
$E(B-V)$ (mag)                              & 0.520 $\pm$ 0.032           & 0.400 $\pm$ 0.048          \\
$E(U-B)$ (mag)                              & 0.276 $\pm$ 0.017           & 0.240 $\pm$ 0.029          \\
$A_{\rm V}$ (mag)                           & 1.612 $\pm$ 0.099           & 1.240 $\pm$ 0.149          \\
$[{\rm Fe/H}]$ (dex)                        & $-0.13 \, \pm$ 0.08         & $-0.01 \, \pm$ 0.06        \\
Age (Myr)                                   & 2400 $\pm$ 200              & 50 $\pm$ 10                \\
Distance modulus (mag)                      & 14.00 $\pm$ 0.12            & 13.56 $\pm$ 0.16           \\
Isochrone distance (pc)                     & 3003 $\pm$ 165              & 2911 $\pm$ 216             \\
$(X, Y, Z)_{\odot}$ (pc)                    & ($-2854$, 925, 126)         & ($-1452$, 2523, 5)         \\
$R_{\rm GC}$ (kpc)                          & 10.89                       & 9.78                       \\
PDMF slope                                  & 1.38 $\pm$ 0.71             & 1.53 $\pm$ 0.39            \\
Total mass ($M/M_{\odot}$)                  & $\sim$181                   & $\sim$118                  \\ 
$V_{\rm \gamma}$ (km/s)                     & $-20.31\, \pm$ 1.86*          &  ---                       \\
$U_{\rm LSR}$ (km/s)                        & 16.62 $\pm$ 2.42            &  ---                       \\
$V_{\rm LSR}$ (km/s)                        & $-26.08 \, \pm$ 1.35        &  ---                       \\
$W_{\rm LSR}$ (km/s)                        & 18.49 $\pm$ 0.69            &  ---                       \\
$S_{_{\rm LSR}}$ (km/s)                     & 36.03 $\pm$ 2.85            &  ---                       \\
$R_{\rm a}$ (pc)                            & 10901 $\pm$ 149             &  ---                       \\
$R_{\rm p}$ (pc)                            & 8893 $\pm$ 34               &  ---                       \\
$Z_{\rm max}$ (pc)                          & 470 $\pm$ 33                &  ---                       \\
$e$                                         & 0.101 $\pm$ 0.005           &  ---                       \\
$P_{\rm orb}$ (Myr)                         & $284 \pm 3$                 &  ---                       \\
Birthplace (kpc)                            & 10.16$^{+0.65}_{-0.71}$     &  ---                       \\
\hline
* \citet{Soubiran18}
        \end{tabular}%
    } %ends the small font selection
    \label{tab:Final_table}%
\end{table}%

%----------------------------------------------------------------------------------------------------------

\section{Summary and Conclusion}

We carried out a detailed study of the two open clusters Berkeley 68 and Stock 20. Analyses were based on CCD {\it UBV} and {\it Gaia} DR3 photometric data together with {\it Gaia} DR3 astrometry. Determination of fundamental and kinematic parameters were made using only stars with a cluster membership probability $P\geq0.5$. The main outcomes of the study are as follows:  

1) RDP fitting to the stellar density distribution determined the cluster limiting radii as $r_{\rm lim}=8'$ ($6.99\pm0.38$ pc) for Berkeley 68 and $r_{\rm lim}=7'.5$ ($6.35\pm0.47$ pc) for Stock 20. These radii are where the stellar densities match those of the surrounding fields about each cluster.  Only stars inside these limiting radii were considered to be potential members of the clusters and subsequently used to derive fundamental cluster parameters.

2) The most probable members of each cluster were selected as follows. Firstly calculation of the membership probabilities was made considering {\it Gaia} DR3 astrometric data and their errors ($\mu_{\alpha}\cos \delta$, $\mu_{\delta}$, $\varpi$). Account was taken of the contamination of the main-sequence by binary stars. This was by fitting the ZAMS to $V\times (B-V)$ CMDs with an additional shift of 0.75 mag in the $V$ band. We identified 198 stars for Berkeley 68 and 51 stars for Stock 20 as probable members (with the cluster membership probability $P\geq0.5$).

3) Mean proper motion components were derived as ($\mu_{\alpha}\cos \delta, \mu_{\delta}) = (2.237\pm 0.007, -1.401\pm 0.005$) mas yr$^{-1}$ for Berkeley 68 and as ($\mu_{\alpha}\cos \delta, \mu_{\delta}) = (-3.215\pm 0.004, -1.172\pm 0.004$) mas yr$^{-1}$ for Stock 20. The mean trigonometric parallaxes were estimated as $\varpi = 0.31 \pm 0.03$ mas for Berkeley 68 and $\varpi = 0.36 \pm 0.03$ mas for Stock 20.  These correspond to distances of $d_{\rm \varpi}=3226\pm 312$ pc and  $d_{\rm \varpi}=2778\pm 232$ pc respectively.

4) Reddening and photometric metallicities were estimated separately for each cluster from {\it UBV} data based TCDs (using most probable main-sequence stars). The reddening and photometric metallicity for Berkeley 68 are $E(B-V)=0.520\pm 0.032$ mag and [Fe/H]$= -0.13\pm 0.08$ dex. The corresponding values for Stock 20 are $E(B-V)=0.400\pm 0.048$ mag and [Fe/H]$= -0.01\pm 0.06$ dex. To derive the distance moduli and ages, we transformed [Fe/H] to mass fraction $z$.  These are $z = 0.012$ for Berkeley 68 and $z = 0.015$ for Stock 20. This is the first time that metallicity has been calculated for either cluster.

5) Distance moduli and ages were derived by fitting {\sc parsec} isochrones to the clusters' {\it UBV} and {\it Gaia} EDR3-based CMDs. The isochrones were selected for metallicities calculated in the immediately previous step. During this step we kept reddening and metallicity as constants. The distance modulus of Berkeley 68 was found to be $\mu_{\rm V}=14.00 \pm 0.12$ mag.  The cluster's distance and age were found to be $d_{\rm iso}=3003 \pm 165$ pc and $t=2400 \pm 200$ Myr, respectively. For Stock 20 the corresponding values are $\mu_{\rm V}=13.56 \pm 0.16$ mag, $d_{\rm iso}=2911 \pm 216$ pc, and $t=50 \pm 10$ Myr. It can be concluded that for both clusters the isochrone fitting distances are compatible with the distances calculated via trigonometric parallaxes ($d_{\rm \varpi}$) from {\it Gaia} DR3. 

6) A radial velocity measurement was only available from the literature for Berkeley 68 but not for Stock 20.  We therefore evaluated galactic orbit parameters and space velocities only for this cluster. We found that the Berkeley 68 belongs to thin-disc component, orbiting outside the solar circle.

7) We concluded from the orbital integration of Berkeley 68 (the previous step) that the cluster was born outside the solar circle at the birth radius $10.16^{+0.65}_{-0.71}$ kpc from the galactic centre and therefore came from a comparatively metal-poor region of the Milky Way.

8) The present day mass function slopes of Berkeley 68 and Stock 20 were estimated as $\Gamma=1.38\pm 0.71$ and $\Gamma=1.53\pm 0.39$, respectively. Both of these values are in good agreement with the value of 1.35 given by \citet{Salpeter55}.

9) Statistical evidence of mass segregation was noticed for Stock 20, with the Kolmogorov-Smirnov test indicating a 97\% confidence level. 

10) Calculation of the relaxation times for each cluster indicates that both clusters are dynamically relaxed: their estimated ages are larger than the relaxation times. 

\software{IRAF \citep{Tody86, Tody93}, PyRAF \citep{Science12}, SExtractor \citep{Bertin96}, Astrometry.net \citep{Lang10}, GALPY \citep{Bovy15}, MWPotential2014 \citep{Bovy15}.}

%---------------------------------------------------------------

\acknowledgments
We thank the anonymous referee for the insightful and constructive suggestions, which significantly improved the paper. This study has been supported in part by the Scientific and Technological Research Council (T\"UB\.ITAK) 120F295. We thank T\"UB\.ITAK for partial support towards using the T100 telescope via project 18CT100-1396. We also thank the on-duty observers and members of the technical staff at the T\"UB\.ITAK National Observatory for their support before and during the observations. TY thanks Tansel Ak for the clusters' observations, as well as Timothy Banks and Sel\c{c}uk Bilir for conversations on cluster analysis. This research has made use of the WEBDA database, operated at the Department of Theoretical Physics and Astrophysics of the Masaryk University. We also made use of VizieR and Simbad databases at CDS, Strasbourg, France. We made use of data from the European Space Agency (ESA) mission \emph{Gaia}\footnote{https://www.cosmos.esa.int/gaia}, processed by the \emph{Gaia} Data Processing and Analysis Consortium (DPAC)\footnote{https://www.cosmos.esa.int/web/gaia/dpac/consortium}. Funding for DPAC has been provided by national institutions, in particular the institutions participating in the \emph{Gaia} Multilateral Agreement. We are grateful for the analysis system IRAF, which was distributed by the National Optical Astronomy Observatory (NOAO). NOAO was operated by the Association of Universities for Research in Astronomy (AURA) under a cooperative agreement with the National Science Foundation. PyRAF is a product of the Space Telescope Science Institute, which is operated by AURA for NASA. We thank the University of Queensland for collaboration software.

%%%%%%%%%%%%%%%%%%%% REFERENCES %%%%%%%%%%%%%%%%%%

\end{document}